\documentclass[12pt,preprint]{aastex}

\slugcomment{Astrophysical Journal}

\shorttitle{Nuclear Starbursts in Seyfert 2 Nuclei}
\shortauthors{Imanishi}

\begin{document}

\title{3--4 $\mu$m Spectroscopy of Seyfert 2 Nuclei to Quantitatively 
Assess the Energetic Importance of Compact Nuclear Starbursts}

\author{Masatoshi Imanishi\altaffilmark{1,2}}
\affil{National Astronomical Observatory, Mitaka, Tokyo 181-8588, Japan}
\email{imanishi@optik.mtk.nao.ac.jp} 

\altaffiltext{1}{Visiting Astronomer at the Infrared Telescope Facility,
which is operated by the University of Hawaii under contract from the 
National Aeronautics and Space Administration.}
                 
\altaffiltext{2}{Visiting Astronomer at the United Kingdom Infrared 
Telescope, which is operated by the Joint Astronomy Centre on behalf 
of the U.K. Particle Physics and Astronomy Research Council.}

\begin{abstract}

We report on 3--4 $\mu$m slit spectroscopy of 13 Seyfert 2 nuclei.  
The 3.3 $\mu$m polycyclic aromatic hydrocarbon (PAH)
emission is used to estimate the magnitudes of compact nuclear starbursts 
(on scales less than a few 100 pc) and to resolve the controversy over their
energetic importance in Seyfert 2 nuclei.
For three selected Seyfert 2 nuclei that have been well studied in the UV,
the magnitudes of the compact nuclear starbursts estimated from the 3.3
$\mu$m PAH emission (with no extinction correction) are in satisfactory
quantitative agreement with those based on the UV after extinction
correction.   
This agreement indicates that the flux attenuation of compact nuclear
starburst emission due to dust extinction is insignificant at 3--4 $\mu$m,
and thus allows us to use the observed 3.3 $\mu$m PAH  luminosity to
estimate the magnitudes of the compact nuclear starbursts in Seyfert 2
nuclei. Based directly on our 3--4 $\mu$m slit spectra, the following
two main conclusions are drawn:
(1) except in one case, the observed nuclear 3--4 $\mu$m emission is dominated
    by AGN and not by starbursts, and
(2) compact nuclear starbursts are detected in 6 out of 13 Seyfert 2
    nuclei, but cannot dominate the energetics of the galactic
    infrared dust emission in the majority of the observed Seyfert 2
    galaxies.
For several sources for which Infrared Space Observatory spectra taken
with larger apertures and/or soft X-ray data are available, these data are
combined with our 3--4 $\mu$m slit spectra, and it is suggested that 
(3) extended (kpc scale) star-formation activity is
    energetically more important than compact nuclear starbursts, and
    contributes significantly to the infrared luminosities of Seyfert 2
    galaxies, although the AGN is still an important contributor to the
    luminosities, and
(4) the bulk of the energetically significant extended star-formation
    activity is of starburst type rather than quiescent normal disk
    star-formation; the extended starbursts are responsible for the
    superwind-driven soft X-ray emission from Seyfert 2 galaxies.
Finally, a correlation between the luminosities of AGNs and compact
nuclear starbursts is implied; more powerful AGNs tend to be related to
more powerful compact nuclear starbursts.

\end{abstract}

\keywords{galaxies: Seyfert --- galaxies: nuclei --- infrared: galaxies}

\section{Introduction}

According to the unification paradigm for Seyfert galaxies, Seyfert 1s,
which show broad optical emission lines, and 2s, which do not, are
intrinsically the same objects, but the nuclei of the latter class are
obscured by dust along our line of sight in dusty molecular tori
\citep{ant93}.  It is widely accepted that the ultimate energy
source of Seyfert {\it nuclei} (although not of their extended host
galaxies) is the release of gravitational energy caused by mass
accretion onto a central supermassive blackhole (so-called AGN
activity).

However, it has recently been argued that strong signatures of
starbursts are detected in the nuclear UV--optical spectra of Seyfert
2 galaxies (Heckman et al. 1997; Gonzalez Delgado et al. 1998;
Storchi-Bergmann et al. 2000; Gonzalez Delgado, Heckman \& Leitherer
2001), and that the intrinsic extinction-corrected luminosities of
compact nuclear starbursts (hereafter `compact' is used to mean a size
scale of less than a few 100 pc) could be comparable to the
luminosities of the AGN \citep{gon98}.  
In contrast,  \citet{iva00}
investigated the CO indices in the near-infrared $K$-band spectra of Seyfert
2 nuclei, and found no evidence for the presence of strong compact nuclear
starbursts.  
Although these authors observed different samples of Seyfert 2 galaxies, 
it is nevertheless striking that they draw such contradictory conclusions
on the energetic role of compact nuclear starbursts in Seyfert 2 nuclei.

According to \citet{cid95}, compact nuclear starbursts are natural
byproducts of AGNs' dusty molecular tori.  In this case, since direct
UV--optical emission from the AGN is highly attenuated in Seyfert 2
nuclei, the less strongly obscured compact nuclear starburst emission will
inevitably make a relatively strong contribution to the observed
UV--optical fluxes in these objects.  Thus, the detection of signatures of
compact nuclear starbursts in the observed UV--optical spectra of
Seyfert 2 nuclei \citep{sto00,gon98,gon01} would not be surprising.  
However, to obtain a deeper understanding
of the nature of Seyfert 2 nuclei, it is certainly more important
to quantify the energetic importance of compact nuclear starbursts
than simply to investigate their presence.  \citet{gon98} estimated
the magnitudes of compact nuclear starbursts, based on their
extinction-corrected UV data, for four UV-bright Seyfert 2
galaxies, while for many other Seyfert 2 galaxies that have been
studied optically, no quantitative discussions of the
absolute magnitudes of compact nuclear starbursts have been made 
\citep{sto00,gon01}. 
Even the UV-based quantitative discussion could contain large
uncertainties, because of the susceptibility of UV emission to dust
extinction; the extinction correction factor for UV data may vary
drastically depending on the assumed amount of dust and
its spatial distribution, and has often been claimed to be
quantitatively uncertain in particular applications
\citep{bak01}.  The uncertainties in the dust
extinction correction are much less in the near-infrared $K$-band data
taken by Ivanov et al.  However, the spectral signatures of starbursts
are weak, so that careful subtraction of old stellar and AGN emission
is required to estimate the magnitudes of compact nuclear starbursts
\citep{iva00}.

At 3--4 $\mu$m, we possess powerful diagnostic tools to distinguish between
starburst and AGN activity, to detect weak starbursts, and to estimate 
the magnitudes of compact nuclear starbursts
in Seyfert 2 galaxies with fewer quantitative uncertainties.
As shown by Imanishi \& Dudley (2000, their figure 1),
\begin{itemize}
\item[(a)] If a Seyfert 2 nucleus is powered by starbursts,
strong polycyclic aromatic hydrocarbon (PAH) emission will be detected
at 3.3 $\mu$m regardless of dust extinction.
\item[(b)] If it is powered by obscured AGN activity, a carbonaceous dust
absorption feature will be detected at 3.4 $\mu$m.
\item[(c)] If it is powered by weakly obscured AGN activity,
the 3--4 $\mu$m spectrum will be nearly featureless.
\item[(d)] If it is a composite of starburst and AGN activity,
the absolute luminosity and the equivalent width of the 3.3 $\mu$m
PAH emission feature will be smaller than in starburst galaxies, so that
these values can be used to quantitatively estimate the energetic
importance of starburst activity.
\end{itemize}
Since the 3.3 $\mu$m PAH emission is intrinsically strong
\citep{moo86,imd00}, 
even the signatures of weak starbursts are detectable;
even if only $\sim$10 \% of the observed nuclear 3--4 $\mu$m
flux originates in starbursts, the 3.3 $\mu$m PAH
emission peak is $\sim$20 \% higher than the continuum level, and thus
is clearly recognizable in normal S/N $\sim$ 15--20 spectra.
Additionally, the effects of dust extinction are smaller at 3--4 $\mu$m
than at shorter wavelengths, which makes the uncertainties in the dust
extinction correction factor much smaller.
This advantage is demonstrated quantitatively below.
Suppose that the dust extinction in starbursts is found to be
A$_{\rm V}$ = 2--4 mag, with an uncertainty of a factor of 2 in
A$_{\rm V}$.
Since the dust extinction in the UV ($\sim$2000 \AA) is $\sim$2.5 $\times$
A$_{\rm V}$ \citep{sav79},
the corresponding dust extinction correction factor in the UV is
100--10000, differing by a factor of 100 depending on the adopted
A$_{\rm V}$.
The difference becomes even larger if we go to shorter-wavelength parts of
the UV region or if the actual dust extinction is larger.
In contrast, since the dust extinction at 3--4 $\mu$m is
$\sim$0.05 $\times$ A$_{\rm V}$ \citep{rie85,lut96}, the extinction
correction factor at 3--4 $\mu$m in the case of A$_{\rm V}$ = 2--4 mag
is 1.1--1.2.
Thus, the extinction correction is negligible at 3--4 $\mu$m, and the
uncertainty in the correction factor is only at the 10\% level.
The high detectability of weak starbursts and the small uncertainties in
dust extinction correction at 3--4 $\mu$m combine to make 3--4 $\mu$m
observations a very powerful tool to quantitatively address the issue of
the energetics of compact nuclear starbursts in Seyfert 2 nuclei.

This paper reports the results of 3--4 $\mu$m spectroscopy of
Seyfert 2 nuclei and their implications for the energetics of these objects. 
Throughout the paper, $H_{0}$ $=$ 75 km s$^{-1}$ Mpc$^{-1}$,
$\Omega_{\rm M}$ = 0.3, and $\Omega_{\rm \Lambda}$ = 0.7 are adopted.

\section{Targets}

The Seyfert 2 nuclei studied by \citet{gon98,gon01} and
\citet{iva00} are selected in order to be able to compare our diagnostic
results at 3--4 $\mu$m directly with those previously obtained in the UV,
optical, and $K$-band. 
Our particular interests are
(1) to assess the quantitative reliability of the extinction correction
    factor based on the UV data \citep{gon98},
(2) to make quantitative the optically based qualitative arguments on the
    magnitudes of compact nuclear starbursts \citep{gon01},
    and
(3) to find weak compact nuclear starbursts whose presence may have been
    missed with $K$-band diagnostics \citep{iva00}.
The observed sources and their properties are summarized in Table~\ref{tbl-1}.

Due to our observing date (Table 2), the observed Seyfert 2
nuclei have right ascensions (R.A.) of between 7 and 16 hr.
All eight Seyfert 2 nuclei in this R.A. range in Table 1
of \citet{gon01} are observed.
\citet{iva00} explicitly presented the strength of the corrected CO
indices, after subtracting AGN emission, in their Table 5.
In total, there are eight Seyfert 2 nuclei (Seyfert type 1.5--2.0) in the
above R.A. range in the Table.
Of these, NGC 5929 was in the sample of \citet{gon01}, leaving seven
independent Seyfert 2 nuclei.
We attempted to observe all seven of these objects during our
observing run, but could not observe UGC 6100.
We started observations on Mrk 270 and Mrk 461, but based on the
first few frames, found that these objects were too faint to obtain spectra
of sufficient quality in a reasonable integration time;  we therefore
discontinued the observations.
Thus, four out of the seven Seyfert 2 nuclei in Table 5 of Ivanov et
al. were observed.
Additionally, a 3--4 $\mu$m spectrum of the very bright Seyfert 2
nucleus NGC 1068 (R.A. = 2 hr), originally presented by \citet{ima97}, is
used because NGC 1068 is also in the list of \citet{gon01}.

In summary, 13 Seyfert 2 nuclei (12 sources with R.A. = 7--16 hr and
NGC 1068) were observed.  For the Seyfert 2 galaxies studied in the
optical by \citet{gon01}, our selection should not be biased with
respect to the presence or absence of compact nuclear starbursts.  For 
the Seyfert 2 galaxies studied in the $K$-band by \citet{iva00}, however,
some bias might be present, since Seyfert 2 galaxies that 
are faint at 3--4 $\mu$m have been excluded.  Nevertheless, our 3--4 $\mu$m
study of these Seyfert 2 nuclei can provide useful information on the issue of
the energetics of compact nuclear starbursts in Seyfert 2 nuclei.

\section{Observations and Data Analysis}

An observing log is tabulated in Table~\ref{tbl-2}.
Details of the observations of NGC 1068 were described by \citet{ima97},
and are not repeated here.
The NSFCAM grism mode \citep{shu94} at IRTF on Mauna Kea,
Hawaii, was used to obtain 3--4 $\mu$m spectra of all the Seyfert 2
galaxies except Mrk 273, Mrk 463, and NGC 1068.
The {\it HKL} grism and {\it L} blocker were used with the 4 pixel
wide slit (= 1$\farcs2$).
The resulting spectral resolution was $\sim$150 at 3.5 $\mu$m.
Mrk 273 and Mrk 463 were observed with the CGS4 (Mountain et al. 1990)
at UKIRT on Mauna Kea, Hawaii.
The 40 l mm$^{-1}$ grating with 2 pixel wide slit
(= 1$\farcs2$) was used.
The resulting spectral resolution was $\sim$750 at 3.5 $\mu$m.
Sky conditions were photometric throughout the observations.
Seeing sizes were 0$\farcs$6--0$\farcs$9 (full width at half maximum) for
both the IRTF and UKIRT observing runs.

Spectra were obtained toward the flux peak at 3--4 $\mu$m.
For sources observed with NSFCAM, the position angles of the slit
were set along the north-south direction.
Consequently, for Mrk 266, which has two nuclei lying on a
southwest to northeast axis, only the southwest Seyfert 2 nucleus
(Mrk 266SW; Mazzarella \& Boroson 1993) was observed.
For Mrk 273 and Mrk 463, which were observed with CGS4,
the position angles were set at 35$^{\circ}$ and 90$^{\circ}$ east of
north, respectively, so as to observe simultaneously emission from both the
double nuclei with a separation of less than 5 arcsec 
\citep{sco00,sur99,sur00}.
A standard telescope nodding technique with a throw of 12 arcsec 
was employed along the slit to subtract background emission.
Since 3--4 $\mu$m emission from Seyfert 2 galaxies is usually dominated by
compact nuclear emission \citep{alo98}, this throw is believed to be
sufficiently large.
Offset guide stars were used whenever available to achieve high telescope
tracking accuracy.
For Mrk 463, the 3--4 $\mu$m emission was dominated by the eastern nucleus
(Mrk 463E) so that the obtained spectrum is almost equivalent to that of
Mrk 463E only.
For Mrk 273, the double nuclei were not clearly resolvable and so emission 
from both nuclei was combined to produce a single spectrum.

F- to G-type standard stars and one B-type star (Table~\ref{tbl-2})
were observed with almost the same airmass as individual Seyfert 2
nuclei, to correct for the transmission of the Earth's atmosphere.
The $L$-band (3.5 $\mu$m) magnitudes of standard stars were estimated
from their $V$-band (0.6 $\mu$m) magnitudes, by adopting the $V-L$
colors appropriate to the stellar types of individual standard stars
(Tokunaga 2000).

Standard data analysis procedures were employed within IRAF
\footnote{
IRAF is distributed by the National Optical Astronomy Observatories,
which are operated by the Association of Universities for Research
in Astronomy, Inc. (AURA), under cooperative agreement with the
National Science Foundation.}.
First, bad pixels were replaced with the interpolated values
of the surrounding pixels.
Next, bias was subtracted from the obtained frames and
the frames were divided by a flat image.
The spectra of the targets and the standard stars were then extracted.
Spectra were extracted by integrating signals over 4--10 arcsec along
the slit (Table~\ref{tbl-3}), depending on the spatial extent of the actual
3--4 $\mu$m signals.
For NSFCAM data, wavelength calibration was performed using
the wavelength-dependent transmission of the Earth's atmosphere, while
an argon lamp was used for the CGS4 data.
The spectra of Seyfert 2 nuclei were divided by those of the standard stars,
and multiplied by the spectra of blackbodies with temperatures
corresponding to individual standard stars (Table~\ref{tbl-2}).
After flux calibration based on the adopted standard star fluxes,
the final spectra were produced.
NSFCAM data at $\sim$3.35 $\mu$m for some sources are contaminated by an
uncorrectable ghost image, and so these data points were removed.

\section{Results}

Flux-calibrated spectra are shown in Figure~\ref{fig1}.  
The spectra of faint sources have been binned by a few spectral
elements.  
The 3--4 $\mu$m spectra of NGC 1068 and Mrk 273, previously presented by
\citet{ima97} and \citet{imd00} respectively, are shown again here. 

The possible energy sources for Seyfert 2 galaxies are AGN,
compact (less than a few 100 pc) nuclear starbursts \citep{hec99}, and
extended (kpc scale) star-formation in the host galaxies.  
The employed slit width of 1$\farcs$2 (except for NGC 1068) corresponds 
to the physical scale of 70--90 pc and 1.1 kpc for the nearest (NGC 3227
and NGC 5033; $z <$ 0.004) and farthest (Mrk 34 and Mrk 463; {\it z} =
0.051) sources, respectively. 
Therefore, except in NGC 3227 and NGC 5033, the bulk of the compact
(less than a few 100 pc) nuclear starburst emission is
covered by our observations.  Although extended (4--10 arcsec)
emission along the slit direction is also included in our spectroscopy,
the fraction of the extended emission inside this thin strip is negligible
compared to the whole.  Therefore, the emission directly investigated based
on our slit spectra is that due to 
AGN and compact nuclear starburst activity, and the contamination
of the emission from extended (kpc scale) star-formation activity is
negligible.  The slit widths employed for the UV, optical, and
$K$-band observations are 1$\farcs$2--1$\farcs$7
\citep{iva00,gon98,gon01}, so that these spectra also reflect the
properties of the AGN and compact nuclear starburst emission.  Thus, 
our 3--4 $\mu$m slit spectra can be directly compared with UV, optical, and  
$K$-band spectra to investigate the nature of AGN and compact nuclear
starbursts.

The bulk of the AGN and compact nuclear starburst emission is, in
principle, detectable in the slit. However, some flux from the compact
emission could be lost in our slit spectra if the telescope tracking
accuracy were insufficient.  3--4 $\mu$m emission from Seyfert 2 galaxies is
generally dominated by compact nuclear emission \citep{alo98}.  
Thus, Table~\ref{tbl-3} compares our
spectro-photometric magnitudes with photometry at 3--4 $\mu$m made with 
apertures of 3--9 arcsec in the literature, in order to estimate what
fraction of compact nuclear 3--4 $\mu$m emission may have been
missed. (We note, however, that time variability of the nuclear 3--4 $\mu$m
emission may affect the comparison.)  In Table~\ref{tbl-3}, our slit
spectrum gives a 
magnitude 1.2 mag fainter than the aperture photometry for NGC 5033. 
This is the closest source in our sample, so that it is quite
plausible that a significant fraction of the 3--4
$\mu$m continuum emission has been missed in this object.  However, since
NGC 5033 is not 
used in the systematic discussion ($\S$ 5), this missing flux for
NGC 5033 will not have a significant effect on our main conclusions.
With the exception of NGC 5033, the magnitude difference
is less than 0.8 mag, or a factor of less than $\sim$2 in all cases.
This agreement within a factor of less than 2 implies that our slit
spectroscopy detects at least half of the compact nuclear emission;
given that aperture photometry may contain a contribution from
extended emission, the actual slit loss for the compact nuclear
emission is even smaller.  Therefore, as far as the compact nuclear
emission is concerned, the absolute 3.3 $\mu$m PAH emission fluxes 
measured with our slit spectra should be certain within a factor of less
than 2.   

To estimate the flux, luminosity, and rest-frame equivalent width of the
3.3 $\mu$m PAH emission, the spectral profile of type-1 sources
\citep{tok91} is adopted as a template of the 3.3 $\mu$m PAH emission. 
This profile can fit well the observed 3.3 $\mu$m PAH emission in some
starbursts with spectra of high spectral resolution and high
signal-to-noise ratios at $\sim$3.3 $\mu$m  
(e.g., NGC 253 and Arp 220 in Imanishi \& Dudley 2000; Mrk 266SW in Fig.1).
The rest-frame peak wavelength of the 3.3 $\mu$m PAH emission is assumed to
be 3.29 $\mu$m, and only the normalization is treated as a free parameter.
The best linear fit for the data points which are affected by neither
emission nor absorption is used as a continuum level. 
Several different continuum levels are adopted, and the uncertainties of
the PAH fluxes (luminosities, equivalent widths) resulting from the
continuum ambiguity are also taken into account, particularly for sources
with small PAH equivalent widths. 
The estimated values are summarized in Table~\ref{tbl-4}.

If dust grains in the obscuring material of Seyfert 2 nuclei are covered 
with an ice mantle, strong H$_{2}$O ice absorption should be detected
\citep{spo00}. This absorption feature has a peak wavelength of 3.08 $\mu$m
and is spectrally broad, extending from 2.8 to 3.6 $\mu$m
(e.g., Smith, Sellgren, \& Tokunaga 1989). The estimates of the 3.3 $\mu$m
PAH fluxes could be highly uncertain if the absorption is strong. However,
most of our spectra cover 3.08 $\mu$m in the rest-frame, and yet no sign of
strong H$_{2}$O ice absorption is found.  
The average spectrum of Seyfert 2 galaxies \citep{cla00} show no
clear H$_{2}$O absorption feature either. 
Since the absorption strength is much weaker at $>$3.25 $\mu$m than at 3.08
$\mu$m \citep{smi89}, possible presence of currently undetectable, weak
H$_{2}$O absorption in Seyfert 2 nuclei is unlikely to affect the estimates 
of the 3.3 $\mu$m PAH emission significantly.  
 
\section{Discussion}

\subsection{What Powers the Observed Compact Nuclear 3--4 $\mu$m Emission?}

The fraction of the observed nuclear 3--4 $\mu$m fluxes in our slit
spectra that originates in starbursts can be estimated from the
rest-frame equivalent widths of the 3.3 $\mu$m PAH emission feature
(EW$_{3.3 \rm PAH}$).
The EW$_{3.3 \rm PAH}$ value decreases as the AGN contribution increases 
($\S$ 1).
The equivalent widths for starburst-dominated galaxies are $\sim$120 nm
\citep{moo86,imd00}.
In Table~\ref{tbl-4}, only Mrk 266SW shows an EW$_{3.3 \rm PAH}$ value
similar to those of starburst galaxies; we can conclude from this that
only the observed nuclear 3--4 $\mu$m flux of Mrk 266SW is dominated by
starbursts.
Mrk 78, Mrk 273, Mrk 477, NGC 3227, and NGC 5135 all show
detectable, moderately strong 3.3 $\mu$m PAH emission features, but
their EW$_{3.3 \rm PAH}$ values are a factor 3--9 smaller than those of
starburst galaxies.
For these five sources, starbursts contribute some fraction, but less than
half, of the observed nuclear 3--4 $\mu$m fluxes, and the bulk of the 
observed 3--4 $\mu$m fluxes originate in AGN activity.
For the remaining seven Seyfert 2 nuclei (IC 3639, Mrk 34, Mrk 463, Mrk 686,
NGC 1068, NGC 5033, and NGC 5929), the EW$_{3.3 \rm PAH}$ is more than a 
factor of 7 smaller than that expected for starburst-dominated
galaxies, indicating that the predominant fraction (larger than 80\%) of
their 3--4 $\mu$m emission comes from AGN activity.
{\it Our first conclusion is that, except in the case of Mrk 266SW,
the observed compact nuclear 3--4 $\mu$m emission is dominated by the AGN
and not by starbursts}. 

This result has implications for infrared studies of AGNs.
$L$- (3.5 $\mu$m) and $M$-band (4.8 $\mu$m) emission from AGNs is usually
dominated by compact emission,
and photometric data at these bands are used to discuss dust obscuration
toward AGNs, on the assumption that this compact emission comes
predominantly from the AGN and not from compact nuclear starbursts
\citep{sim98,ima01}.
Our results provide supporting evidence for this assumption.

\subsection{The 3.4 $\mu$m Dust Absorption Feature}

The 3.4 $\mu$m carbonaceous dust absorption feature is detected in Mrk 463
and NGC 1068.
The observed 3.4 $\mu$m absorption optical depths are
$\tau_{3.4}$(observed) $=$ 0.042$\pm$0.005 and 0.12$\pm$0.01 for Mrk 463
and NGC 1068, respectively.
In AGNs, the 3--4 $\mu$m continuum emission is dominated by dust
in thermal equilibrium with a temperature of 800--1000K, close to the
dust sublimation temperature, located at the innermost part of the dusty
torus \citep{sim98,alo98}.
The dust extinction toward the 3--4 $\mu$m continuum emitting region
is thus almost the same as that toward the AGN itself.
Assuming a Galactic dust model ($\tau_{3.4}$/A$_{\rm V}$ $=$ 0.004--0.007;
Pendleton et al. 1994), the dust extinction toward the AGNs is estimated to
be 6--11 and 17--30 mag for Mrk 463 and NGC 1068, respectively.
Mrk 463 shows a broad emission component in its near-infrared hydrogen
recombination lines \citep{good94,vei97} so that dust obscuration
toward the AGN has been
suggested to be relatively modest in this object.
Our result is consistent with this suggestion.

For the remaining 11 Seyfert 2 nuclei, no detectable 3.4 $\mu$m absorption
feature is found.
The 3.4 $\mu$m absorption feature may be suppressed if the bulk of the dust
grains in the obscuring material is covered with an ice mantle
\citep{men01}. However, no strong 3.08 $\mu$m ice absorption feature is 
found in their 3--4 $\mu$m spectra.  
The dust absorption feature is smeared out if less obscured starburst
emission contributes significantly to the observed nuclear 3--4 $\mu$m 
flux \citep{ima00}. The six sources with detectable PAH emission (Mrk 78,
Mrk 266SW, Mrk 273, Mrk 477, NGC 3227, and NGC 5135) could be this case. 

For the five PAH-undetected Seyfert 2 nuclei (IC 3639, Mrk 34, Mrk 686, 
NGC 5033, and NGC 5929), the 3.4 $\mu$m absorption feature could remain
undetectable if dust obscuration toward these AGNs is weak ($\S$ 1). 
The R-value, defined as
\begin{eqnarray}
R \equiv log \frac{\nu_{3.5 \mu m} F(\nu_{3.5 \mu m})}{\nu_{25 \mu m}
F(\nu_{25 \mu m})}
\end{eqnarray}
is a good measure for dust extinction toward AGNs (Murayama, Mouri, \& 
Taniguchi 2000). 
It becomes smaller with higher dust extinction toward AGNs, and 
the boundary value between Seyfert 1 and 2 nuclei is estimated to be 
R = $-$0.6 \citep{mura00}. 
Based on the 3.5 $\mu$m (Table \ref{tbl-3}) and {\it IRAS} 25 $\mu$m
(Table \ref{tbl-1}) fluxes, the R-values for IC 3639, Mrk 34, Mrk 686, 
NGC 5033, and NGC 5929, are estimated to be $-$1.2, $-$0.8, $-$0.3, 
$-$0.6, and $-$1.2, respectively. 
The R-values of Mrk 686 and NGC 5033 are in the range of Seyfert 1
nuclei, and thus weak dust obscuration is indicated, explaining the
non-detection of the 3.4 $\mu$m absorption.   
However, IC 3639, Mrk 34, and NGC 5929 show small R-values. The
non-detection could be caused by the effects of time variability or
contamination from extended star-formation to the observed 25 $\mu$m flux
taken with large apertures \citep{alo01}.

\subsection{The Magnitudes of Compact Nuclear Starbursts as Estimated from the
3.3 $\mu$m PAH Emission Fluxes}

The bulk of the UV--optical emission from the AGNs in Seyfert 2 galaxies
and starbursts \citep{cal00} is absorbed by dust and re-emitted as dust
thermal emission in the infrared.
Infrared (8--1000 $\mu$m) luminosities are thus used to estimate
the magnitudes of AGN and starburst activity throughout this paper.
Infrared (8--1000 $\mu$m) luminosities are obviously better for evaluating
these kinds of activity than far-infrared (40--500 $\mu$m)
luminosities because not all activity shows dust emission peaking in the
far-infrared.

In starburst-dominated galaxies, the 3.3 $\mu$m PAH to far-infrared
(40--500 $\mu$m) luminosity ratios (L$_{3.3 \rm PAH}$/L$_{\rm FIR}$) are
$\sim$1 $\times$ 10$^{-3}$ \citep{mouri90}.
The 3.3 $\mu$m PAH to infrared (8--1000 $\mu$m) luminosity ratio (L$_{3.3
\rm PAH}$/L$_{\rm IR}$) is tentatively assumed to be also
1 $\times$ 10$^{-3}$ for the compact nuclear starbursts in Seyfert 2 nuclei.
The scatter of the PAH-to-infrared luminosity ratio is likely to be a
factor of 2--3 toward both higher and lower values around a typical value 
\citep{fis00}.

For IC 3639, Mrk 477, and NGC 5135, \citet{gon98} explicitly estimated
the extinction-corrected luminosities of compact nuclear starbursts, based
on their UV data.
The respective luminosities are 1.1 $\times$ 10$^{43}$, 1.4 $\times$
10$^{44}$, and 4.2 $\times$ 10$^{43}$ ergs s$^{-1}$.
The infrared luminosities of the compact nuclear starbursts estimated from
the observed (extinction-uncorrected) 3.3 $\mu$m PAH emission, based on the
assumption of L$_{3.3 \rm PAH}$/L$_{\rm IR}$ $\sim$ 1 $\times$ 10$^{-3}$,
are $<$1.2 $\times$ 10$^{43}$, 1.5 $\times$ 10$^{44}$, and
5.8 $\times$ 10$^{43}$ ergs s$^{-1}$, respectively,
for IC 3639, Mrk 477, and NGC 5135.
It is noteworthy that the compact nuclear starburst luminosities estimated
from UV data after extinction correction and from the
3.3 $\mu$m PAH emission with {\it no} extinction correction are in
satisfactory agreement for all three sources.
This agreement indicates that
(1) discussions of the magnitudes of compact nuclear starbursts based on UV
data after extinction correction are quantitatively reliable,
(2) the extinction correction factor is not significant at 3--4 $\mu$m,
and
(3) the assumption of L$_{3.3 \rm PAH}$/L$_{\rm IR}$ $\sim$
1 $\times$ 10$^{-3}$ is reasonable.
We have been concerned by the possibility that, if compact nuclear starbursts
close to the central AGNs are directly exposed to X-ray emission from AGNs, 
PAH emission from the compact nuclear starbursts might be suppressed due to
the destruction of PAHs \citep{voit92}. 
However, the agreement suggests that this is not the case.
If the bulk of the compact nuclear starbursts in Seyfert 2 nuclei occur at the
outer reaches of the dusty tori around AGNs \citep{hec97},
PAHs are shielded from the energetic radiation from the AGNs and thus
are not destroyed, explaining the strong PAH emission from compact nuclear
starbursts in Seyfert 2 nuclei.
Based on these findings, the magnitudes of compact 
nuclear starbursts can be estimated from the observed 3.3 $\mu$m PAH
emission luminosities with some confidence. 

If compact nuclear starbursts energetically dominated the whole galactic
infrared dust emission luminosities of Seyfert 2 galaxies, the observed
L$_{3.3 \rm PAH}$/L$_{\rm IR}$ ratios measured from our slit spectra
should be similar to those of starburst galaxies
($\sim$1 $\times$ 10$^{-3}$).
As shown in Table~\ref{tbl-4}, however, none of the observed Seyfert 2
galaxies show such a large value of L$_{3.3 \rm PAH}$/L$_{\rm IR}$.
The differences are by a factor of 3 (Mrk 477) to larger than 10, and 
in most cases much larger than a factor of 2--3, the possible flux loss in
our slit spectra and the intrinsic scatter of the L$_{3.3 \rm PAH}$/L$_{\rm
IR}$ ratios for starbursts. 
Therefore, {\it our second conclusion is that compact nuclear starbursts
cannot be the dominant energy source of the whole galactic infrared dust
emission luminosities in the majority of the observed Seyfert 2 galaxies}.

\subsection{Comparison with Optical and $K$-band Diagnostic Results for
Individual Sources}

For the remaining ten Seyfert 2 galaxies other than IC 3639, Mrk 477, and
NGC 5135, our quantitative estimates of the magnitudes of the compact
nuclear starbursts are compared with qualitative arguments based on the 
optical and $K$-band data \citep{gon01,iva00}. 

\citet{gon01} found signatures of compact nuclear starbursts (young stars)
in the optical spectrum of Mrk 273.  For Mrk 273, 3.3 $\mu$m
PAH emission is detected; EW$_{3.3 \rm PAH}$ and L$_{3.3 \rm PAH}$/L$_{\rm IR}$
are a factor of $\sim$3 and $\sim$10 smaller, respectively, than those
of starburst galaxies.  
The L$_{3.3 \rm PAH}$/L$_{\rm IR}$ ratio indicates that the compact nuclear
starbursts explain $\sim$1/10 of the total galactic infrared luminosity.  
Thus, compact nuclear starbursts are certainly present, supporting the
optical results.

For Mrk 78 and NGC 1068, \citet{gon01} found no signatures of compact 
nuclear starbursts, and detected only intermediate age stars in their
optical spectra.
For NGC 1068, only stringent upper limits are found for the
EW$_{3.3 \rm PAH}$ and L$_{3.3 \rm PAH}$/L$_{\rm IR}$, both of which
are more than an order of magnitude smaller than those of starburst
galaxies.
For Mrk 78, however, 3.3 $\mu$m PAH emission is marginally detected.

For Mrk 463, \citet{gon01} claimed that compact nuclear starbursts might be
present, subject to further confirmation.
No detectable 3.3 $\mu$m PAH emission is found. 
The upper limits for both EW$_{3.3 \rm PAH}$ and 
L$_{3.3 \rm PAH}$/L$_{\rm IR}$ are more than an order of magnitude 
smaller than starburst galaxies. 
Compact nuclear starbursts in Mrk 463 are energetically insignificant, if
they exist at all.

For Mrk 34, \citet{gon01} found no signs of compact nuclear starbursts in
the optical spectrum, with only old population stars being detected.
No 3.3 $\mu$m PAH emission is detected either, supporting the conclusions
based on the optical data.

Both \citet{gon01} and \citet{iva00} studied NGC 5929 and
found no evidence for strong compact nuclear starbursts.
No clear 3.3 $\mu$m PAH emission is detected.

The remaining sources, Mrk 266SW, Mrk 686, NGC 3227, and NGC 5033, were
studied only by \citet{iva00}.  They found that no strong compact nuclear
starbursts are required in terms of the CO indices after correction for
AGN emission.  No 3.3 $\mu$m PAH emission is detected in Mrk 686 and NGC
5033, supporting their argument.  However, in Mrk 266SW, the nuclear
3--4 $\mu$m emission must be dominated by starbursts, although
these detected compact nuclear starbursts can explain only $\sim$10\% of
the total galactic infrared luminosity.  3.3 $\mu$m PAH emission was also
clearly detected in NGC 3227, and compact nuclear starbursts can explain
$\sim$20\% of the whole galactic infrared luminosity.  
In Mrk 266SW and NGC 3227, compact nuclear starbursts are certainly
present, and they explain a non-negligible fraction of the total galactic
infrared luminosities. 
The signatures of compact nuclear starbursts may have been missed during
the complicated procedures required to find their weak signatures in the
$K$-band \citep{iva00}.

\subsection{The Energetic Importance of AGN and Extended Star-Formation
Activity}

Since the observed nuclear 3--4 $\mu$m emission is dominated by AGN
activity ($\S$ 5.1), it would be expected that AGN activity also
contributes much more than compact nuclear starbursts to the net galactic
infrared (8--1000 $\mu$m) luminosities of Seyfert 2 galaxies.
However, the AGN-driven infrared luminosities of these objects are
difficult to estimate because they are highly dependent on the spatial
distribution of dust in the dusty torus, which is poorly constrained
observationally.  Thus, we estimate the energetic importance of extended
star-formation in the host galaxies, which is the remaining probable power
source of Seyfert 2 galaxies other than compact nuclear starbursts and
AGNs, by using the PAH emission.  
PAH emission can be produced both with starbursts and normal quiescent
star-formation in a similar way \citep{hel00}.

To investigate the magnitude of extended (kpc scale)
star-formation activity, {\it Infrared Space Observatory (ISO)} spectra
taken with large apertures are useful.  \citet{gh00} summarized the {\it ISO}
results; they found that (1) the 7.7 $\mu$m PAH to far-infrared
luminosity ratios in Seyfert galaxies are similar to those of
starburst galaxies, and that (2) the bulk of the PAH emission is
spatially extended.  Therefore, they argued that the bulk of the
far-infrared emission from Seyfert galaxies originates in extended
star-formation activity.

Of the 13 Seyfert 2 galaxies, \citet{cla00} have presented {\it ISO}
2.5--11 $\mu$m spectra taken with 24 $\times$ 24 arcsec$^{2}$ apertures,
and quote PAH fluxes for four sources (Mrk 266, NGC 3227, NGC 5033, and NGC
5929).  For these sources, the energetic importance of extended
star-formation activity can be investigated by comparing the measured PAH
fluxes in our slit spectra with those in the {\it ISO} spectra.  

The effects of dust extinction are similar at 3--8 $\mu$m \citep{lut96}.
The 7.7 $\mu$m PAH emission is the strongest PAH emission feature. 
However, its flux estimates in the {\it ISO} spectra may be highly
uncertain \citep{cla00,lau00}, due to the presence of strong, spectrally
broad 9.7 $\mu$m silicate dust absorption feature and insufficient
wavelength coverage longward of these emission and absorption features, 
which make a continuum determination difficult \citep{dud99}. 
Quantitatively reliable flux estimates of the 3.3 $\mu$m PAH emission 
in the {\it ISO} spectra are also difficult due to the scatter in the data
points at 3--4 $\mu$m (Clavel et al. 2000, their figure 8).
Consequently, the 6.2 $\mu$m PAH emission, which is isolated and
moderately strong, is most suitable to investigate the extended
star-formation activity based on the {\it ISO} spectra \citep{fis00}. 

The 6.2 $\mu$m PAH to infrared luminosity ratios
(L$_{6.2 \rm PAH}$/L$_{\rm IR}$) for starbursts are estimated to be 
 $\sim$ 6 $\times$ 10$^{-3}$, with a scatter of a factor of 2--3 toward
both higher and lower values \citep{fis00}, where it is assumed that 
L$_{\rm IR}$ $\sim$ L$_{\rm FIR}$ for starbursts.
The respective L$_{6.2 \rm PAH}$/L$_{\rm IR}$ ratios are 4 $\times$
10$^{-3}$, 4 $\times$ 10$^{-3}$, 3 $\times$ 10 $^{-3}$, and 1 $\times$
10$^{-3}$, for Mrk 266, NGC 3227, NGC 5033, and NGC 5929 \citep{cla00}, 
roughly half of (Mrk 266, NGC 3227, and NGC 5033) or more than six times
smaller than (NGC 5929) the typical value for systems dominated by
starbursts.  
Since the observed L$_{3.3 \rm PAH}$/L$_{\rm IR}$ ratios indicate that 
the compact nuclear starbursts energetically fall short of the total
infrared luminosities by a factor of larger than 6 for these four sources
(Table~\ref{tbl-4}), it can be said that extended star-formation (but not
compact nuclear starbursts) contributes significantly to the infrared
luminosities of these Seyfert 2 galaxies.
This statement was made by \citet{gh00} for Seyfert galaxies as a
whole based on the 7.7 $\mu$m PAH emission.
Based on the 6.2 $\mu$m PAH emission, here, it has been confirmed that this
is true also for the four individual Seyfert 2 galaxies, and probably for
Seyfert 2 galaxies as a whole.
The individual L$_{6.2 \rm PAH}$/L$_{\rm IR}$ ratios for Mrk 266,
NGC 3227, and NGC 5033 are within the scattered range of the ratios for
starbursts. However, the overall trend of lower values implies that AGN
activity also makes an important contribution to the infrared luminosities
of Seyfert 2 galaxies.  

\subsection{Is Extended Star-Formation Quiescent or of Starburst Type?}

While both starbursts and quiescent disk star-formation in normal galaxies
can produce PAH emission ($\S$ 5.5), strong soft X-ray emission driven by
superwind is observed only if the star-formation rate per unit area exceeds
a certain threshold
(10$^{-1}$ M$_{\odot}$ yr$^{-1}$ kpc$^{-2}$; Heckman 2000).
Starbursts surpass this threshold, while quiescent normal disk
star-formation does not \citep{ken98}.
These different characteristics can be used to understand the properties
of energetically significant extended star-formation activity.

Of the 13 Seyfert 2 galaxies, \citet{lev01} have estimated superwind-driven
soft
X-ray luminosities for eight sources (IC 3639, Mrk 78, Mrk 266, Mrk 273,
Mrk 463, Mrk 477, NGC 1068, and NGC
5135), and found that, as a whole, their soft X-ray to far-infrared
luminosity ratios are as high as those of starburst galaxies.
Therefore, starburst activity is a significant contributor to the far-infrared
emission and also to the infrared dust emission.
Since compact nuclear starbursts detected in our slit
spectra were found to be energetically insignificant ($\S$ 5.3), the
energetically significant starbursts must be extended. 
{\it Our third conclusion is that the bulk of the energetically significant
extended (kpc scale) star-formation activity is of starburst-type and
not quiescent normal disk star-formation.
It is the extended (kpc scale) starbursts, rather than the compact (less
than a few 100 pc) nuclear starbursts, that are responsible for the
superwind-driven soft X-ray emission.}
If starbursts are inevitable phenomena in Seyfert 2 galaxies,
their extended starburst activity is energetically more important than
their compact nuclear starbursts.
This is actually the case for the four Seyfert 2 galaxies studied in detail
in the UV \citep{gon98} and for the famous, well-studied Seyfert 2 galaxy
NGC 1068 \citep{lef01}.

\subsection{Do More Powerful AGNs Have More Powerful Compact Nuclear
Starbursts?}

We test the hypothesis that more powerful AGNs might be related to
more powerful compact nuclear starbursts \citep{gon98}. 
Figure~\ref{fig2} compares the 12 $\mu$m luminosity with the 3.3 $\mu$m PAH
emission luminosity.   
The total galactic 12 $\mu$m luminosity is regarded as a measure of 
AGN power \citep{gon01}. 
The 3.3 $\mu$m PAH luminosities, measured with our slit spectroscopy,
reflect the magnitudes of compact nuclear starbursts.  The picture of
\citet{gon98} would be supported if we were to find a positive correlation
between 12 $\mu$m luminosity and 3.3 $\mu$m PAH emission luminosity.

In Fig.~\ref{fig2}, this trend appears to be present, although the
scatter is moderately large.
We apply the generalized Kendall's rank correlation statistic \citep{iso86}
to the data points in Fig.~\ref{fig2}, and estimate the probability that a
correlation is not present to be 0.11, by using the software available at
the web page: http://www.astro.psu.edu/statcodes/.
Thus, provided that the 12 $\mu$m luminosity is a good measure of AGN
power, {\it it is found that the luminosities of AGNs and compact nuclear
starbursts in Seyfert 2 galaxies are correlated, and more powerful AGNs
tend to contain more powerful compact nuclear starbursts.}

\section{Summary}

We performed 3--4 $\mu$m spectroscopy of 13 Seyfert 2 nuclei
that were previously studied in the UV, optical, and near-infrared $K$-band.
Making use of the 3.3 $\mu$m PAH emission feature detected in our slit
spectra, the following results were obtained.

\begin{enumerate}
\item Our PAH-based, extinction-uncorrected estimates 
      of the luminosities of compact (less than a few 100 pc) nuclear
      starbursts were found to agree well with those based on
      extinction-corrected UV data in 
      three selected Seyfert 2 galaxies (IC 3639, Mrk 477, and NGC 5135).
      This agreement indicates that compact nuclear starburst
      emission is not significantly attenuated due to dust extinction at
      3--4 $\mu$m, so that the observed 3.3 $\mu$m PAH emission luminosity
      is a powerful diagnostic of the magnitudes of compact nuclear
      starbursts. 
\item For the remaining ten Seyfert 2 galaxies,
      our PAH-based diagnostic results were compared with
      qualitative arguments based on the optical and $K$-band spectra.
      For Mrk 273, signs of compact nuclear starbursts were
      found, supporting the qualitative optically based argument.
      For Mrk 463, possible compact nuclear starbursts implied from the
      optical data were not confirmed at 3--4 $\mu$m.
      For Mrk 78 and NGC 1068, both of which show signatures of only
      intermediate age stars in the optical, 3.3 $\mu$m PAH emission was
      marginally detected in Mrk 78. Optical and/or $K$-band spectra show no
      evidence for compact nuclear starbursts in Mrk 34, Mrk 266SW, Mrk
      686, NGC 3227, NGC 5033, and NGC 5929. At 3--4 $\mu$m, signatures of
      compact nuclear starbursts were found in Mrk 266SW and NGC 3227.
\item Quantitative estimates of the magnitudes of nuclear starbursts
      were made based on the observed 3.3 $\mu$m PAH emission luminosities
      for the 13 Seyfert 2 galaxies,
      and found that (1) nuclear 3--4 $\mu$m emission is dominated by AGN
      activity rather than by starburst activity in all
      Seyfert 2 galaxies but one (Mrk 266SW), and (2) compact nuclear
      starbursts can explain only a small fraction ($<$1/10--1/3) 
      of the infrared dust emission luminosities.
\item The 3.4 $\mu$m carbonaceous dust absorption feature is detected only
      in the two Seyfert 2 nuclei, Mrk 463 and NGC 1068.
\item When our observations were combined with large-aperture {\it ISO}
      spectra for the four sources (Mrk 266, NGC 3227, NGC 5033, and NGC 
      5929), it was confirmed that extended (kpc scale)
      star-formation activity is energetically more important than compact
      nuclear starbursts, and contributes significantly to the infrared
      dust emission luminosities of Seyfert 2 galaxies.  However, the AGNs
      are likely to be still important contributors to the luminosities.
\item Making use of soft X-ray data for the eight sources (IC 3639, Mrk 78,
      Mrk 266, Mrk 273, Mrk 463, Mrk 477, NGC 1068, and NGC 5135), it 
      was suggested that the energetically significant extended
      star-formation activity is of starburst-type and not quiescent normal
      disk star-formation. 
      These extended starbursts are responsible for the superwind-driven
      soft X-ray emission from Seyfert 2 galaxies.
\item We find evidence supporting the scenario in which more powerful
      AGNs are related to more powerful compact nuclear starbursts,
      provided that 12 $\mu$m luminosity is a good measure of AGN power.
\end{enumerate}

\acknowledgments

We thank S. K. Leggett, J. Davies, T. Wold, and T. Carroll for their
support during the UKIRT observing runs, and J. Rayner and B. Golisch
for their support before and during the IRTF observing run.
We are grateful to Drs. K. Aoki, T. Nakajima, and Y. P. Wang for their
useful comments on this manuscript, and Dr. T. T. Takeuchi for invaluable
discussions about statistical tests.
The anonymous referee and the scientific editor, Dr. S. Willner, gave  
invaluable comments, which improved this paper significantly.
Spectra of Mrk 273 and Mrk 463 were obtained while MI was at the University
of Hawaii.
This research has made use of the NASA/IPAC Extragalactic Database
(NED) which is operated by the Jet Propulsion Laboratory, California
Institute of Technology, under contract with the National Aeronautics
and Space Administration.

\clearpage

\begin{deluxetable}{lcrrrrccc}
\tablecaption{Summary of Seyfert 2 nuclei. \label{tbl-1}}
\tablewidth{0pt}
\tablehead{
\colhead{Object} & \colhead{Redshift}   & 
\colhead{f$_{\rm 12}$}   & 
\colhead{f$_{\rm 25}$}   & 
\colhead{f$_{\rm 60}$}   & 
\colhead{f$_{\rm 100}$}  & 
\colhead{log L$_{\rm FIR}$} & 
\colhead{log L$_{\rm IR}$} & 
\colhead{Remarks} \\
\colhead{} & \colhead{}   & \colhead{[Jy]}   & \colhead{[Jy]} & 
\colhead{[Jy]}  & \colhead{[Jy]} & \colhead{[ergs s$^{-1}$]} & 
\colhead{[ergs s$^{-1}$]} & \colhead{} \\
\colhead{(1)} & \colhead{(2)} & \colhead{(3)} & \colhead{(4)} & 
\colhead{(5)} & \colhead{(6)} & \colhead{(7)} & \colhead{(8)} & \colhead{(9)}  
}
\startdata
IC 3639  & 0.011 & 0.64 & 2.26 & 7.52 & 10.7 & 44.25 & 44.33 & G \\ 
Mrk 34   & 0.051 & 0.07 & 0.46 & 0.81 & 0.80 & 44.59 & 44.78 & G \\
Mrk 78   & 0.037 & 0.13 & 0.56 & 1.11 & 1.13 & 44.44 & 44.63 & G \\
Mrk 266  & 0.028 & 0.23 & 0.98 & 7.34 & 11.1 & 45.07 & 45.03 & I \\
Mrk 273  & 0.038 & 0.24 & 2.28 & 21.7 & 21.4 & 45.75 & 45.69 & G \\ 
Mrk 463  & 0.051 & 0.51 & 1.58 & 2.18 & 1.92 & 45.00 & 45.34 & G \\  
Mrk 477  & 0.038 & 0.13 & 0.51 & 1.31 & 1.85 & 44.58 & 44.70 & G \\ 
Mrk 686  & 0.014 & $<$0.11 & 0.13 & 0.57 & 1.79 & 43.49 & 43.43--43.56 & I \\ 
NGC 1068 & 0.004 & 39.70 & 85.04 & 176.2 & 224.0 & 44.72 & 44.96 & G \\
NGC 3227 & 0.004 & 0.67 & 1.76 & 7.83 & 17.6 & 43.46 & 43.49 & I \\ 
NGC 5033 & 0.003 & 0.95 & 1.15 & 13.8 & 43.9 & 43.54 & 43.48 & I \\
NGC 5135 & 0.014 & 0.64 & 2.40 & 16.9 & 28.6 & 44.84 & 44.81 & G \\ 
NGC 5929 & 0.008 & 0.43 & 1.62 & 9.14 & 13.7 & 44.06 & 44.06 & G, I \\ 
\enddata
       
\tablecomments{
Column (1): Object. Column (2): Redshift.
Column (3)--(6): f$_{12}$, f$_{25}$, f$_{60}$, and f$_{100}$ are 
{\it IRAS FSC}
fluxes at 12$\mu$m, 25$\mu$m, 60$\mu$m, and 100$\mu$m, respectively.
Column (7): Logarithm of far-infrared (40--500 $\mu$m) luminosity
in ergs s$^{-1}$ calculated with
$L_{\rm FIR} = 1.4 \times 2.1 \times 10^{39} \times$ $D$(Mpc)$^{2}$
$\times (2.58 \times f_{60} + f_{100}$) [ergs s$^{-1}$]
\citep{sam96}.
Column (8): Logarithm of infrared (8$-$1000 $\mu$m) luminosity
in ergs s$^{-1}$ calculated with
$L_{\rm IR} = 2.1 \times 10^{39} \times$ D(Mpc)$^{2}$
$\times$ (13.48 $\times$ $f_{12}$ + 5.16 $\times$ $f_{25}$ +
$2.58 \times f_{60} + f_{100}$) [ergs s$^{-1}$]
\citep{sam96}.
Column (9): G: sources studied by Gonzalez Delgado et al. (2001).
I: sources studied by Ivanov et al. (2000).}

\end{deluxetable}

\clearpage

\begin{deluxetable}{llccccc}
\tablecaption{Observing Log. \label{tbl-2}}
\tablewidth{0pt}
\tablehead{
\colhead{} & \colhead{Date}   & \colhead{Integration}   & 
\multicolumn{4}{c}{Standard Star} \\
\cline{4-7} \\
\colhead{Object} & \colhead{(UT)}   & \colhead{Time (sec)}  & 
\colhead{Star Name} & \colhead{$L$-mag}  & \colhead{Type} & 
\colhead{T$_{\rm eff}$ (K)} 
}
\startdata
IC 3639 & 2001 Apr 8  & 1920 & HR 5212 & 4.8 & F7V & 6240 \\
Mrk 34  & 2001 Apr 9  & 2520 & HR 4112 & 3.4 & F8V & 6200 \\
Mrk 78  & 2001 Apr 9  & 1728 & HR 3028 & 4.8 & F6V & 6400 \\
Mrk 266SW & 2001 Apr 9  & 2160 & HR 4767 & 4.8 & F8V--G0V & 6000 \\
Mrk 273 & 2000 Feb 20 & 2560 & HR 4761 & 4.8 & F6--8V & 6200 \\
Mrk 463 & 2000 Jun 13 &  800 & HR 5243 & 4.9 & F6V & 6400 \\ 
Mrk 477 & 2001 Apr 9  & 2592 & HR 5581 & 4.3 & F7V & 6240 \\ 
Mrk 686 & 2001 Apr 9  & 2400 & HR 5423 & 4.7 & G5V & 5700 \\ 
NGC 1068 & 1995 Nov 15 & 640 & HR 8781 & 2.6 & B9V & 10700 \\
NGC 3227 & 2001 Apr 8 & 1050 & HR 3650 & 4.6 & G9V & 5400 \\
NGC 5033 & 2001 Apr 8 & 2280 & HR 5423 & 4.7 & G5V & 5700 \\
NGC 5135 & 2001 Apr 8 & 2160 & HR 5212 & 4.8 & F7V & 6240 \\
NGC 5929 & 2001 Apr 9 & 2064 & HR 5581 & 4.3 & F7V & 6240 \\ 
\enddata
\end{deluxetable}

\clearpage

\begin{deluxetable}{lccc}
\tablecaption{Our Spectro-photometric $L$-band Magnitudes and 
Comparisons with Aperture Photometry at $L$ in the Literature. \label{tbl-3}}
\tablewidth{0pt}
\tablehead{
\colhead{} & \multicolumn{2}{c}{Magnitude} &
\colhead{} \\
\cline{2-3} \\
\colhead{Object} & \colhead{(our data)} & \colhead{(literature)} 
& \colhead{Reference} \\
\colhead{(1)} & \colhead{(2)} & \colhead{(3)} & \colhead{(4)} 
}
\startdata
IC 3639 & 10.2 (1$\farcs$2 $\times$ 5$''$) & 10.4 (6$''$) & G\\
Mrk 34  & 11.3 (1$\farcs$2 $\times$ 4$''$) & 11.1 (8$\farcs$5) & R \\
Mrk 78  & 11.0 (1$\farcs$2 $\times$ 5$''$) & 11.0 (5$\farcs$9) & R \\
Mrk 266SW & 12.1 (1$\farcs$2 $\times$ 5$''$) & 
 11.5 (8$\farcs$6) & I \\
Mrk 273 & 10.5 (1$\farcs$2 $\times$ 10$''$) & 10.5 (5$''$)& Z \\
Mrk 463 &  8.2 (1$\farcs$2 $\times$ 7$''$) & 8.2 (4$''$) & M\\
Mrk 477 & 11.1 (1$\farcs$2 $\times$ 4$''$) & 11.8 (6$''$) & L \\
Mrk 686 & 11.0 (1$\farcs$2 $\times$ 5$''$) & 11.2 (5$\farcs$4) & I \\
NGC 1068 & 4.8 (3$\farcs$8 $\times$ 3$\farcs$8) & 4.8 (3$''$) & MA \\
NGC 3227 & 9.5 (1$\farcs$2 $\times$ 4$''$) & 9.0 (8$\farcs$6) & I \\
NGC 5033 & 10.7 (1$\farcs$2 $\times$ 5$''$) & 9.5 (8$\farcs$6) & I \\
NGC 5135 & 10.1 (1$\farcs$2 $\times$ 6$''$) & 9.7 (6$''$) & G  \\
NGC 5929 & 11.6 (1$\farcs$2 $\times$ 5$''$) & 10.8 (5$\farcs$4) & I \\
\enddata

\small

\tablecomments{
Column (1): Object.
Column (2): Our spectro-photometric magnitude at $L$.
Column (3): Aperture photometric $L$-band magnitude from the literature.
Column (4): References for the aperture photometry.
G: Glass and Moorwood (1985), I: Ivanov et al. (2001),
L: Lawrence et al. (1985), M: Mazzarella et al. (1991),
MA: Marco \& Alloin (2000), R: Rieke (1978),
Z: Zhou, Wynn-Williams, \& Sanders (1993).
}

\end{deluxetable}

\clearpage

\begin{deluxetable}{lcccc}
\tabletypesize{\small}
\tablecaption{The Properties of the 3.3 $\mu$m PAH Emission Feature. 
\label{tbl-4}}
\tablewidth{0pt}
\tablehead{
\colhead{} & \colhead{f$_{3.3 \rm PAH}$} & \colhead{L$_{3.3 \rm PAH}$}  &
\colhead{L$_{3.3 \rm PAH}$/L$_{\rm IR}$} & \colhead{rest EW$_{3.3 \rm PAH}$} \\
\colhead{Object} & \colhead{($\times$ 10$^{-14}$ ergs s$^{-1}$ cm$^{-2}$)} & 
\colhead{($\times$ 10$^{39}$ ergs s$^{-1}$)} & 
\colhead{($\times$ 10$^{-3}$)} 
& \colhead{($\times$ 100 nm)} \\
\colhead{(1)} & \colhead{(2)} & \colhead{(3)} & \colhead{(4)} 
& \colhead{(5)}  
}
\startdata
IC 3639 & $<$4.9 & $<$11.3 & $<$5.2 $\times$ 10$^{-2}$ & 
$<$8.5 $\times$ 10$^{-2}$       \\
Mrk 34   & $<$0.33      & $<$17.5     & $<$2.9 $\times$ 10$^{-2}$ 
         & $<$1.6 $\times$ 10$^{-2}$        \\
Mrk 78   & 3.9$\pm$1.0  & 108$\pm$27 & 2.5 $\times$ 10$^{-1}$ 
         & 1.5$\pm$0.4 $\times$ 10$^{-1}$ \\
Mrk 266SW & 8.5$\pm$0.5 & 133$\pm$7 & 1.2 $\times$ 10$^{-1}$ 
         & 8.7$\pm$0.5 $\times$ 10$^{-1}$ \\
Mrk 273  & 14.5$\pm$0.7 & 422$\pm$19 & 8.7 $\times$ 10$^{-2}$ 
         & 3.5$\pm$0.2 $\times$ 10$^{-1}$ \\
Mrk 463  & $<$3.0  & $<$159 & $<$7.4 $\times$ 10$^{-2}$ 
         & $<$8.1 $\times$ 10$^{-3}$ \\
Mrk 477  & 5.1$\pm$0.7  & 151$\pm$21 & 3.0 $\times$ 10$^{-1}$ 
         & 2.2$\pm$0.3 $\times$ 10$^{-1}$  \\
Mrk 686  & $<$3.3      & $<$12      & $<$4.6 $\times$ 10$^{-1}$       
& $<$1.2 $\times$ 10$^{-1}$        \\
NGC 1068 & $<$86.7      & $<$26.7     & $<$2.9 $\times$ 10$^{-2}$      
& $<$1.2 $\times$ 10$^{-2}$     \\
NGC 3227 & 16.6$\pm$1.3 & 5.1$\pm$0.4 & 1.7 $\times$ 10$^{-1}$ 
         & 1.4$\pm$0.1 $\times$ 10$^{-1}$  \\
NGC 5033 & $<$0         & $<$0        & $<$0        & $<$0        \\
NGC 5135 & 15.3$\pm$2.8 & 58$\pm$11 & 9.1 $\times$ 10$^{-2}$ 
         & 2.4$\pm$0.4 $\times$ 10$^{-1}$ \\
NGC 5929 & $<$2.5  & $<$3.0  & $<$2.7 $\times$ 10$^{-2}$ 
         & $<$1.4 $\times$ 10$^{-1}$    \\
\enddata

\tablecomments{Column (1): Object.
Column (2): Observed 3.3 $\mu$m PAH flux.
Column (3): Observed 3.3 $\mu$m PAH luminosity.
Column (4): Observed 3.3 $\mu$m PAH to infrared luminosity ratio
            in units of 10$^{-3}$, the typical value for starburst
            galaxies.
Column (5): Rest frame equivalent width of the 3.3 $\mu$m PAH emission
            in units of 100 nm, approximately the typical value for
            starburst galaxies ($\sim$120 nm).}
\end{deluxetable}

\clearpage

\begin{figure}
\plottwo{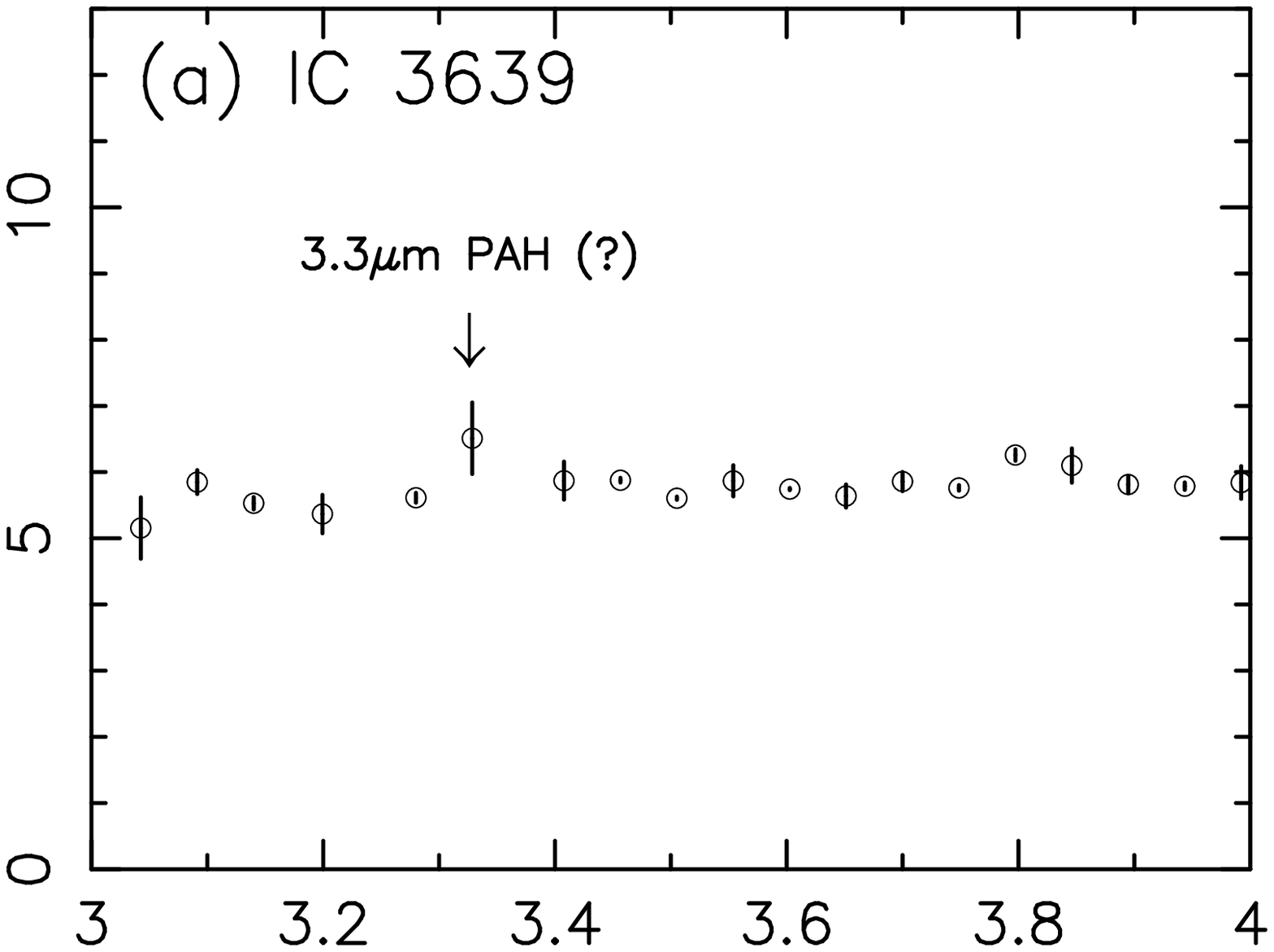}{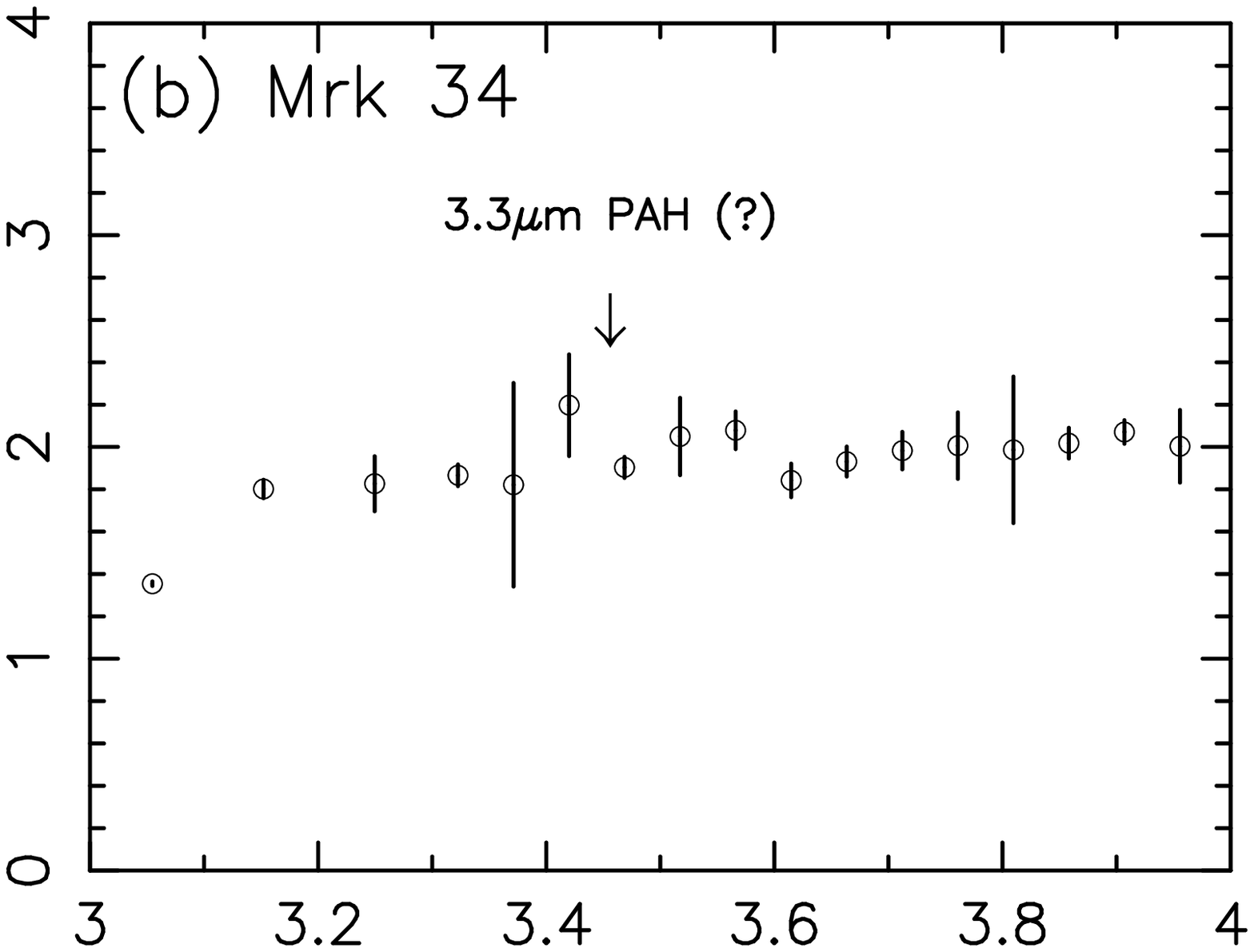} 
\end{figure}
\begin{figure}
\plottwo{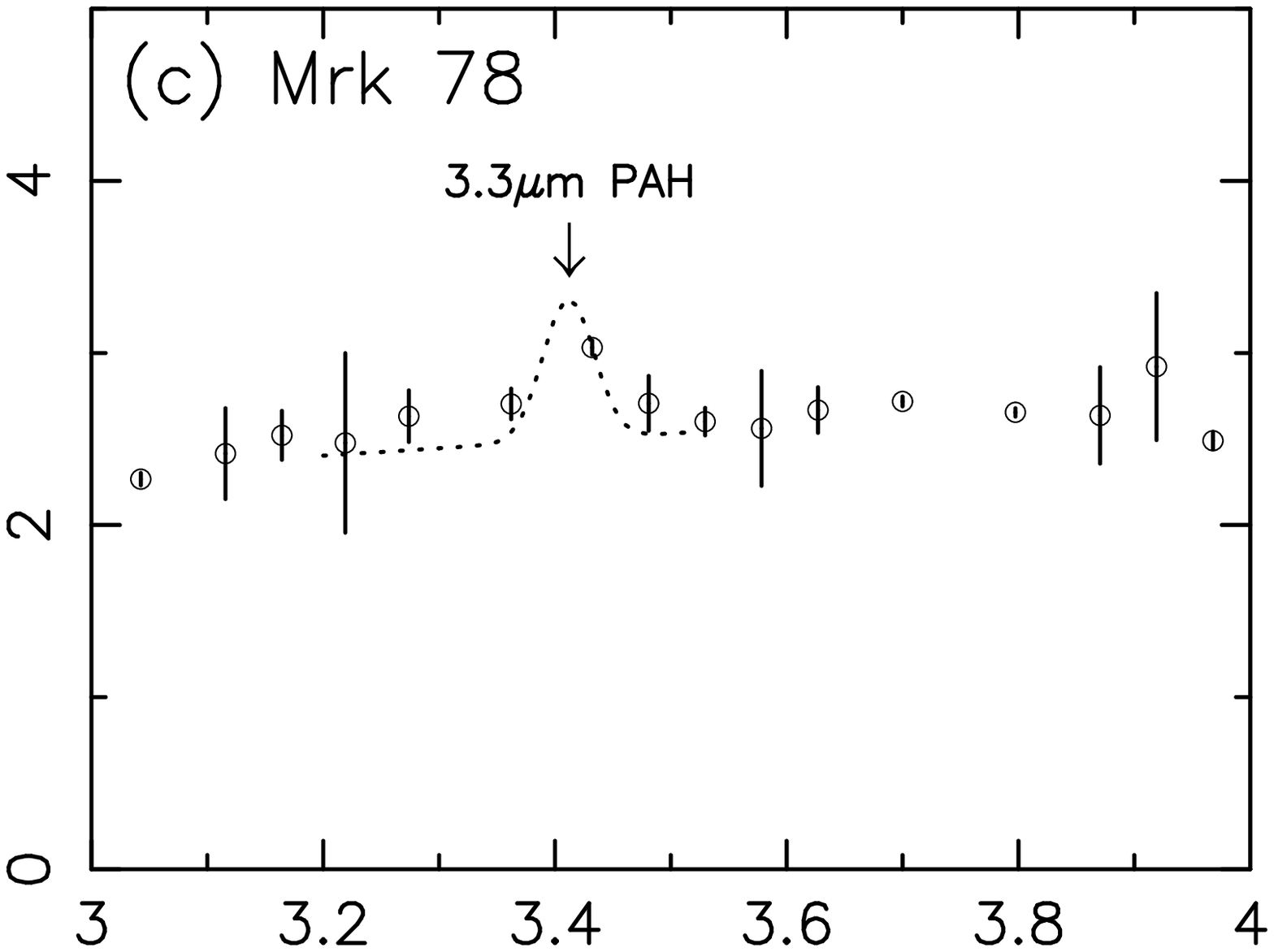}{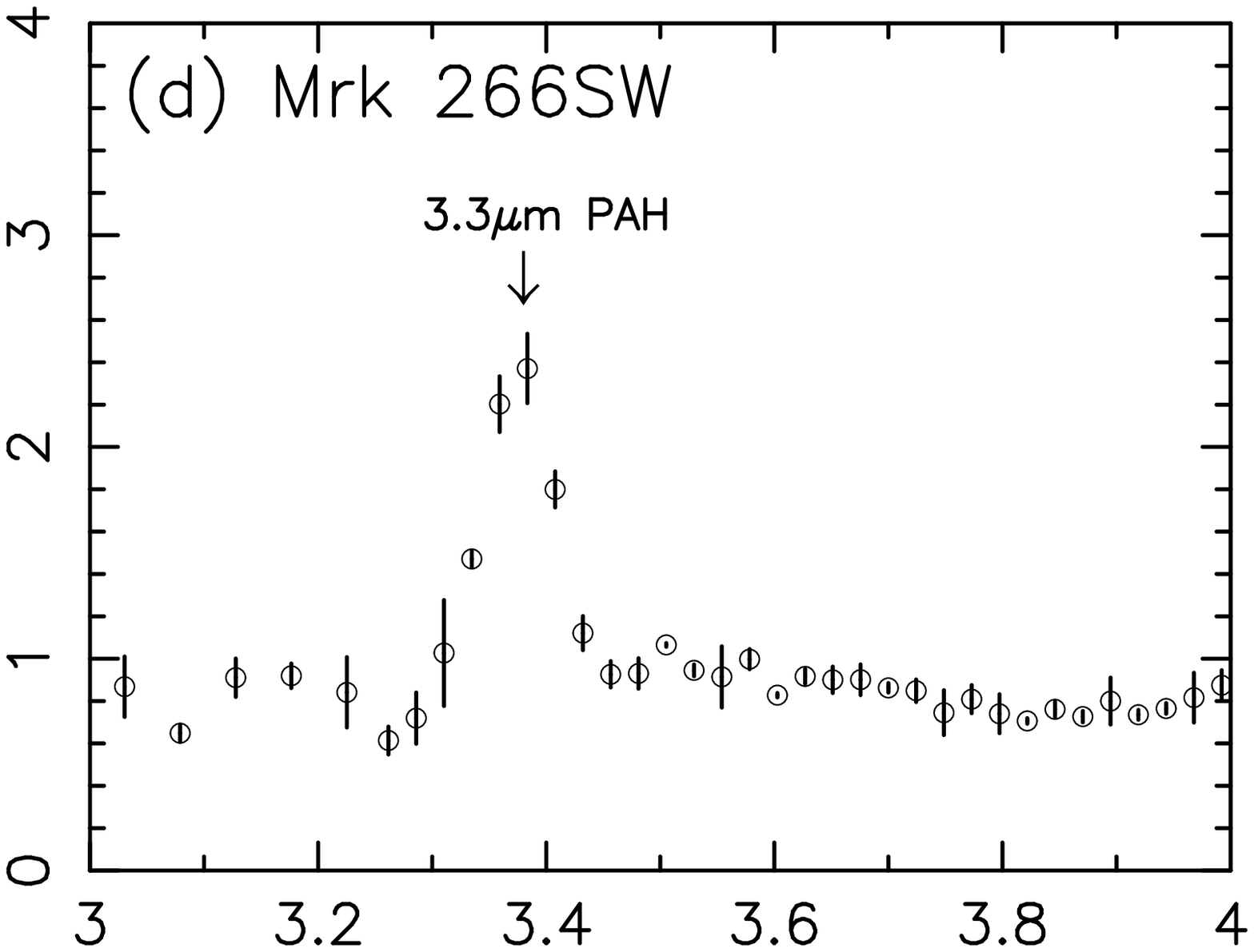}
\end{figure}
\begin{figure}
\plottwo{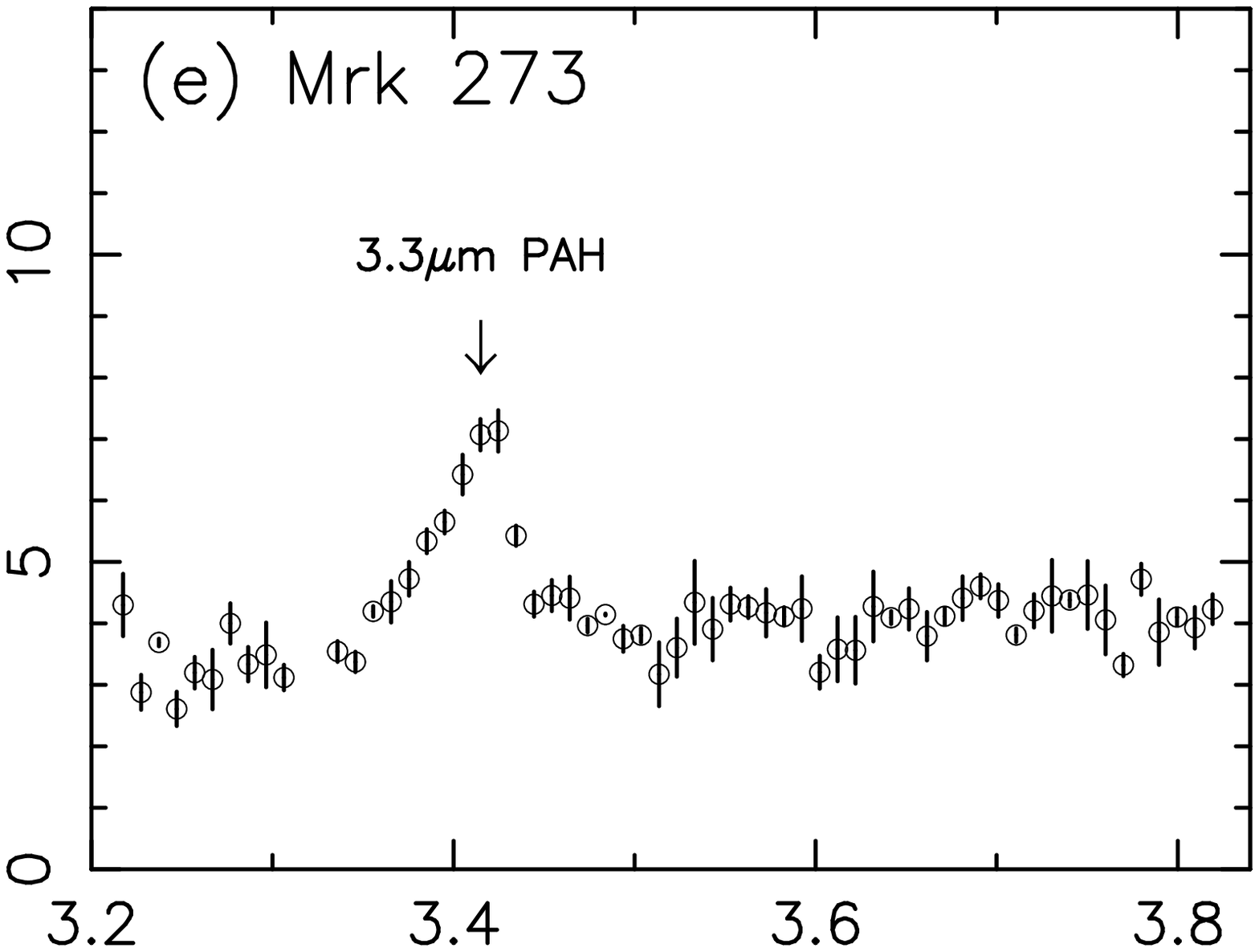}{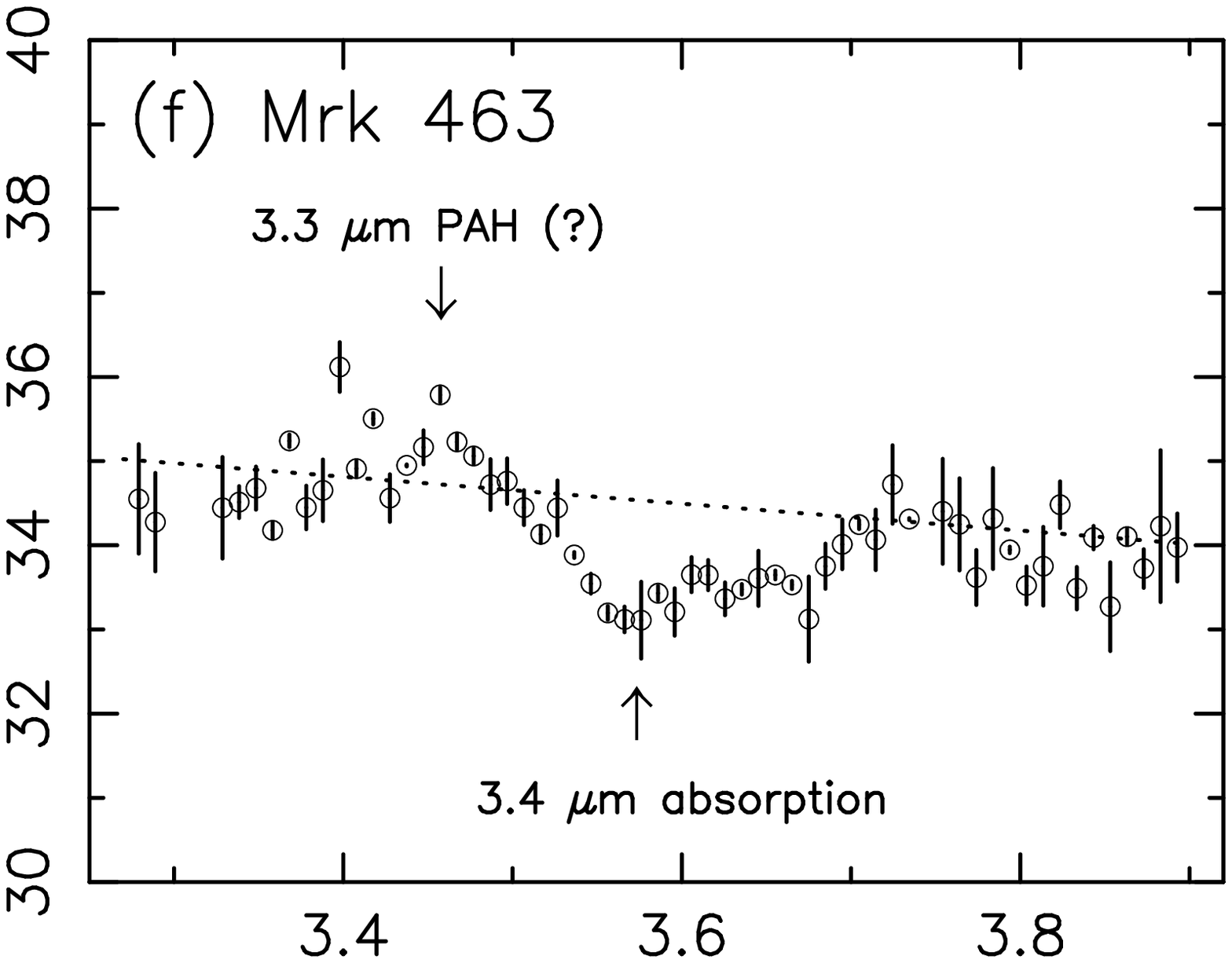}
\end{figure}

\clearpage

\begin{figure}
\plottwo{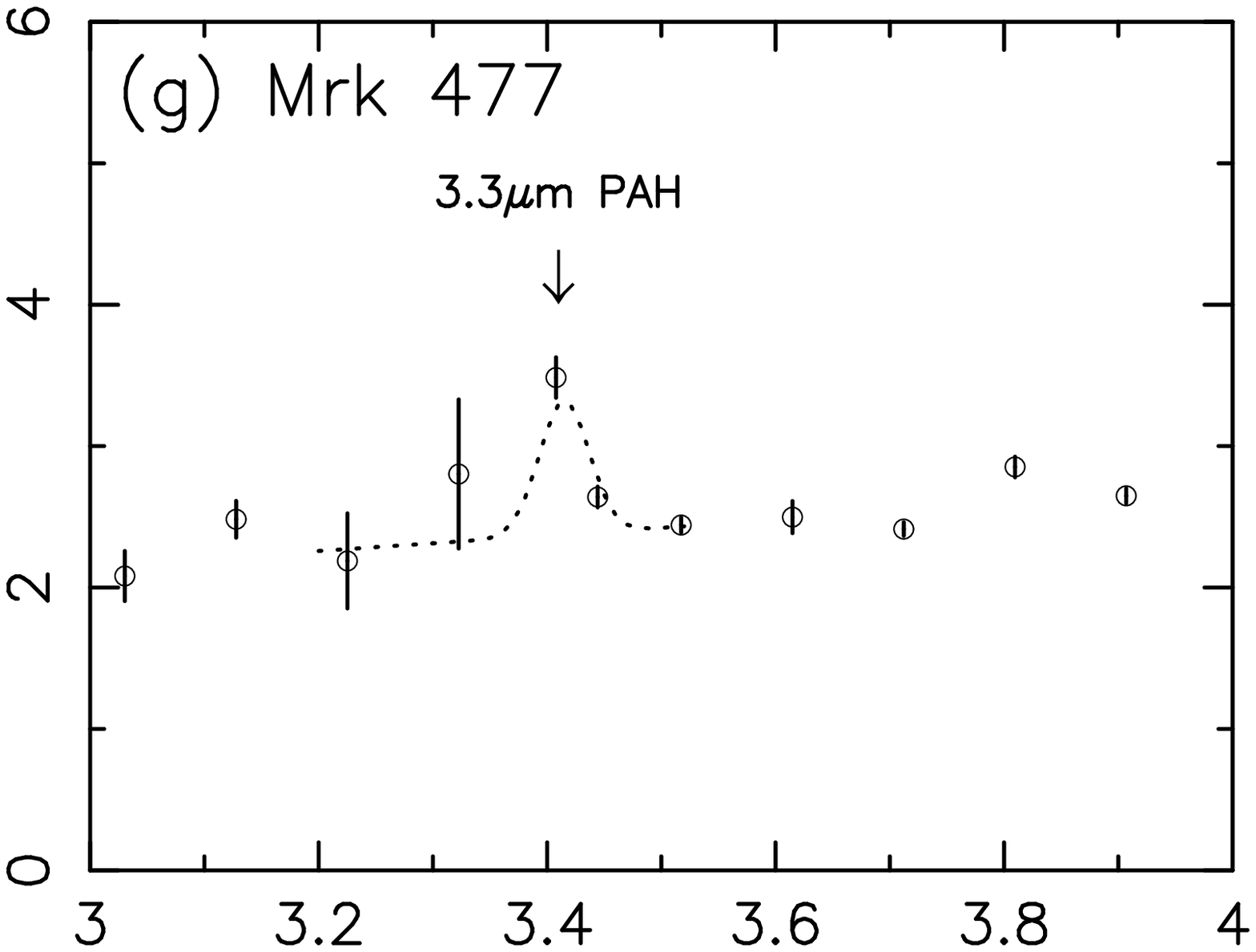}{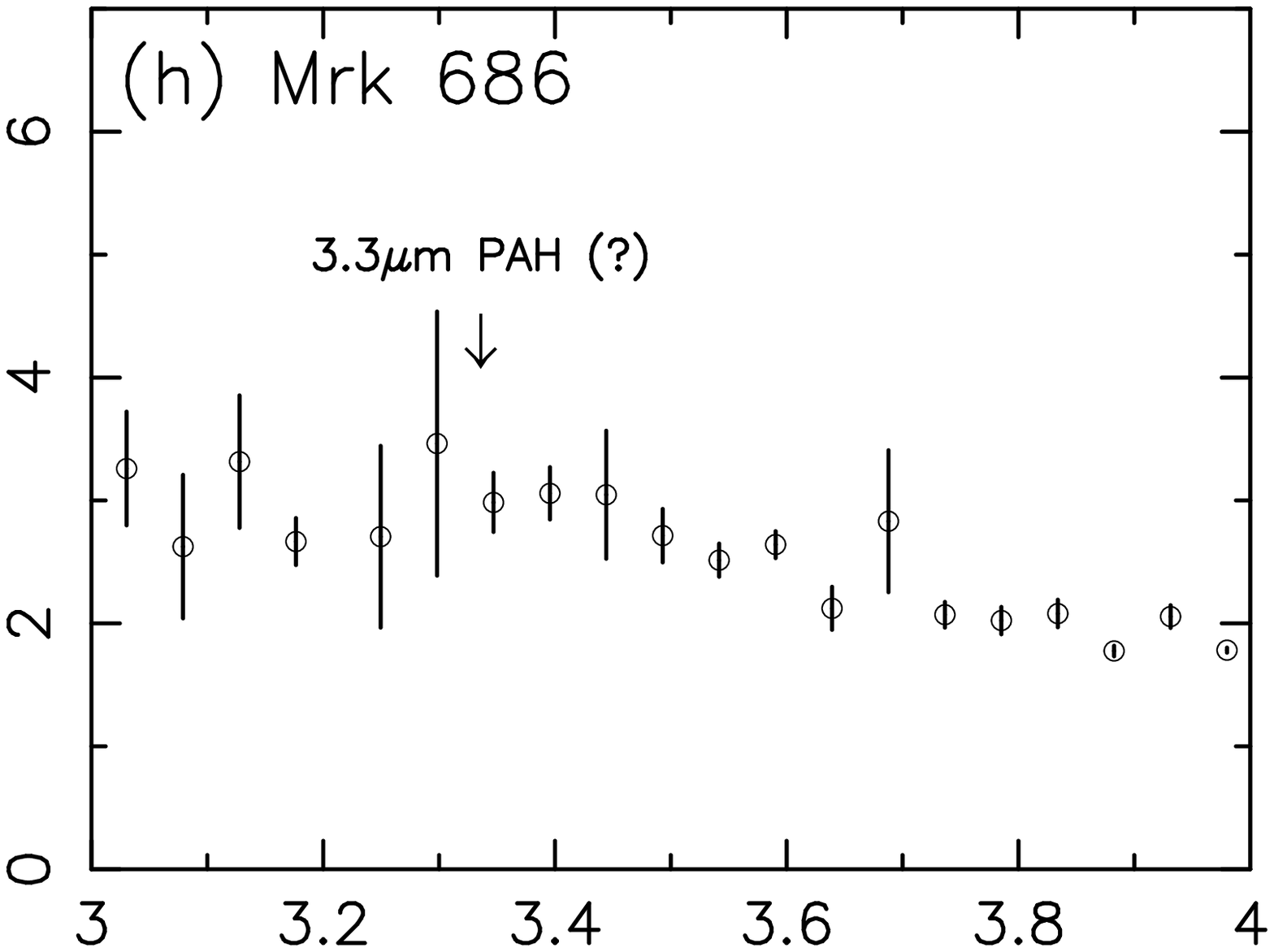}
\end{figure}
\begin{figure}
\plottwo{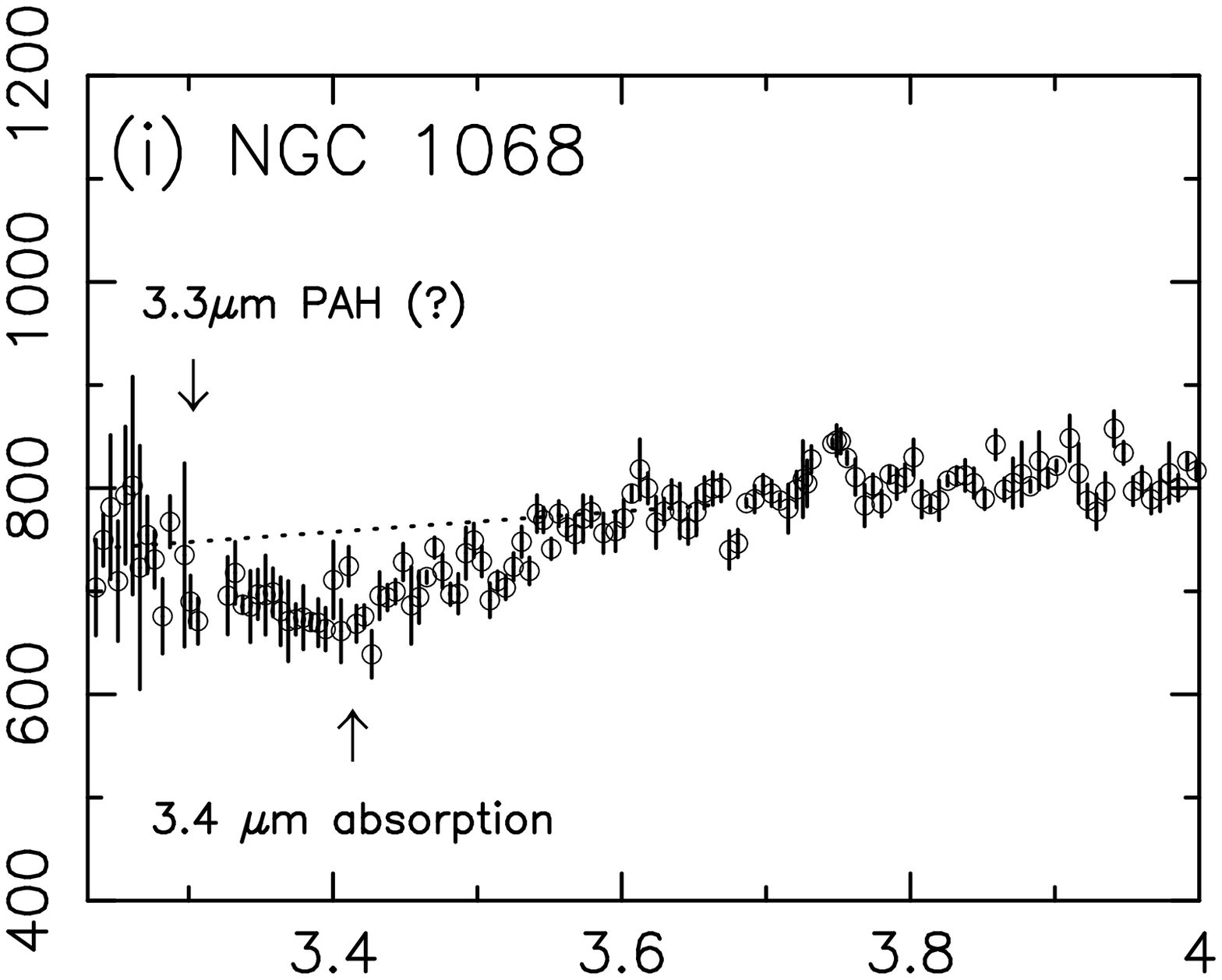}{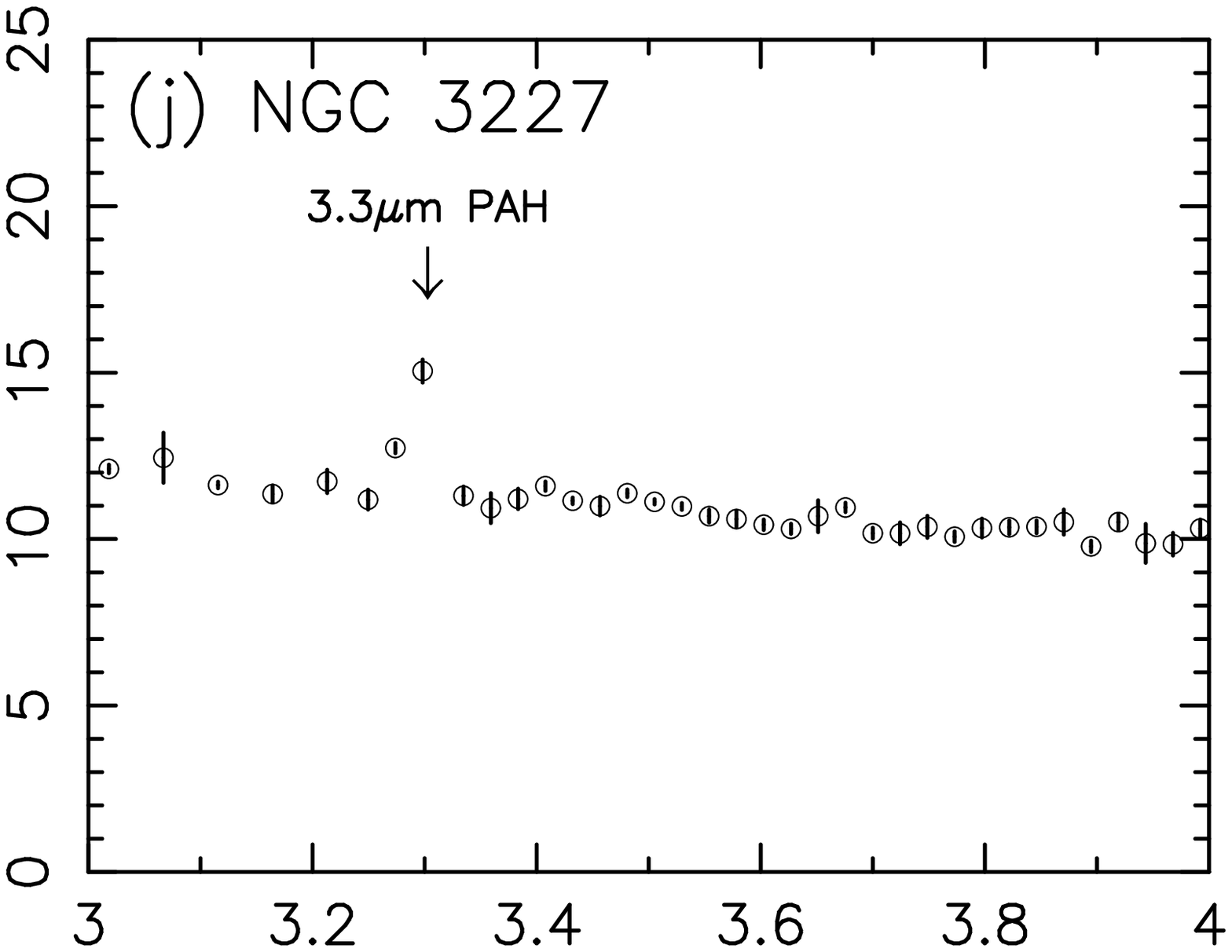}
\end{figure}
\begin{figure}
\plottwo{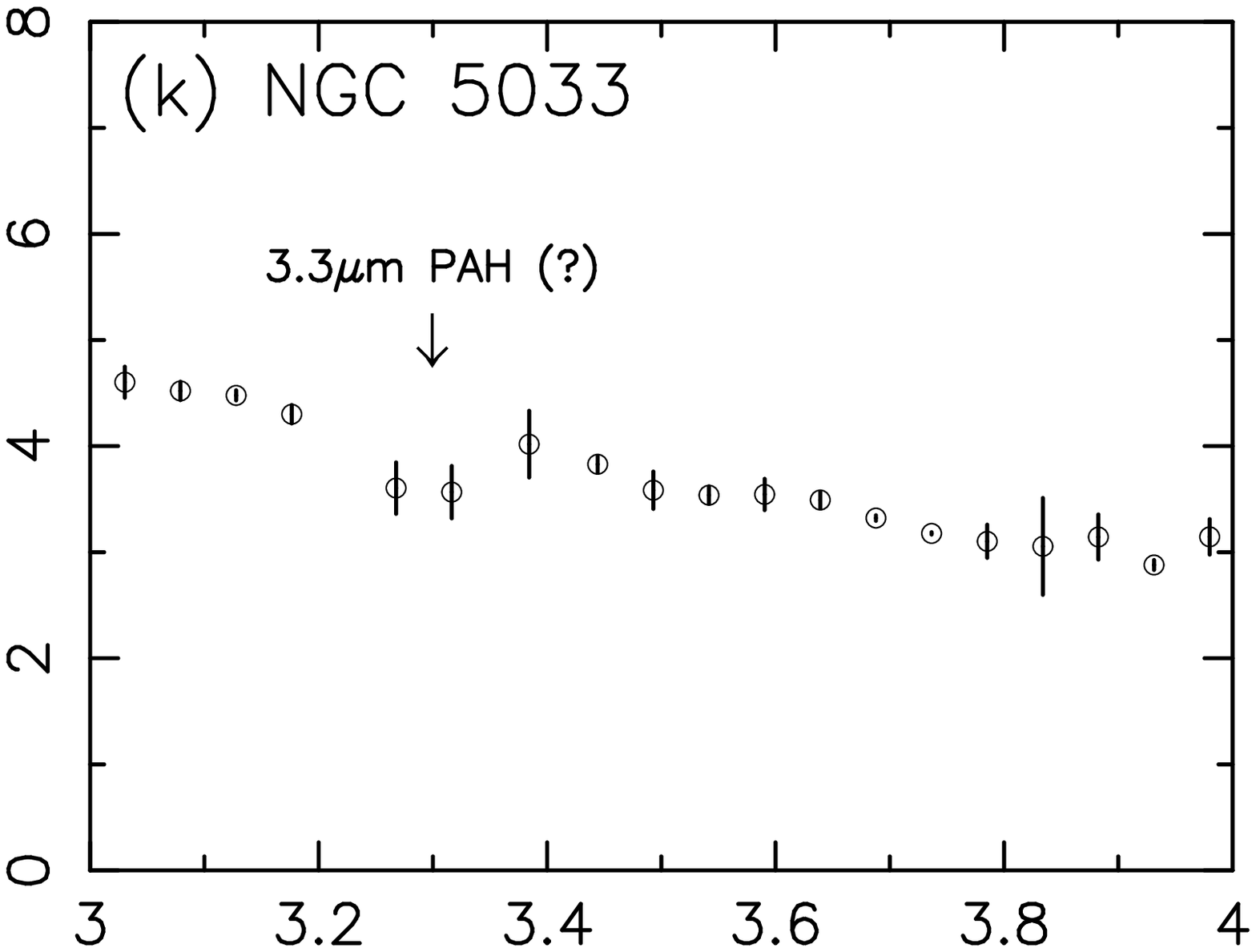}{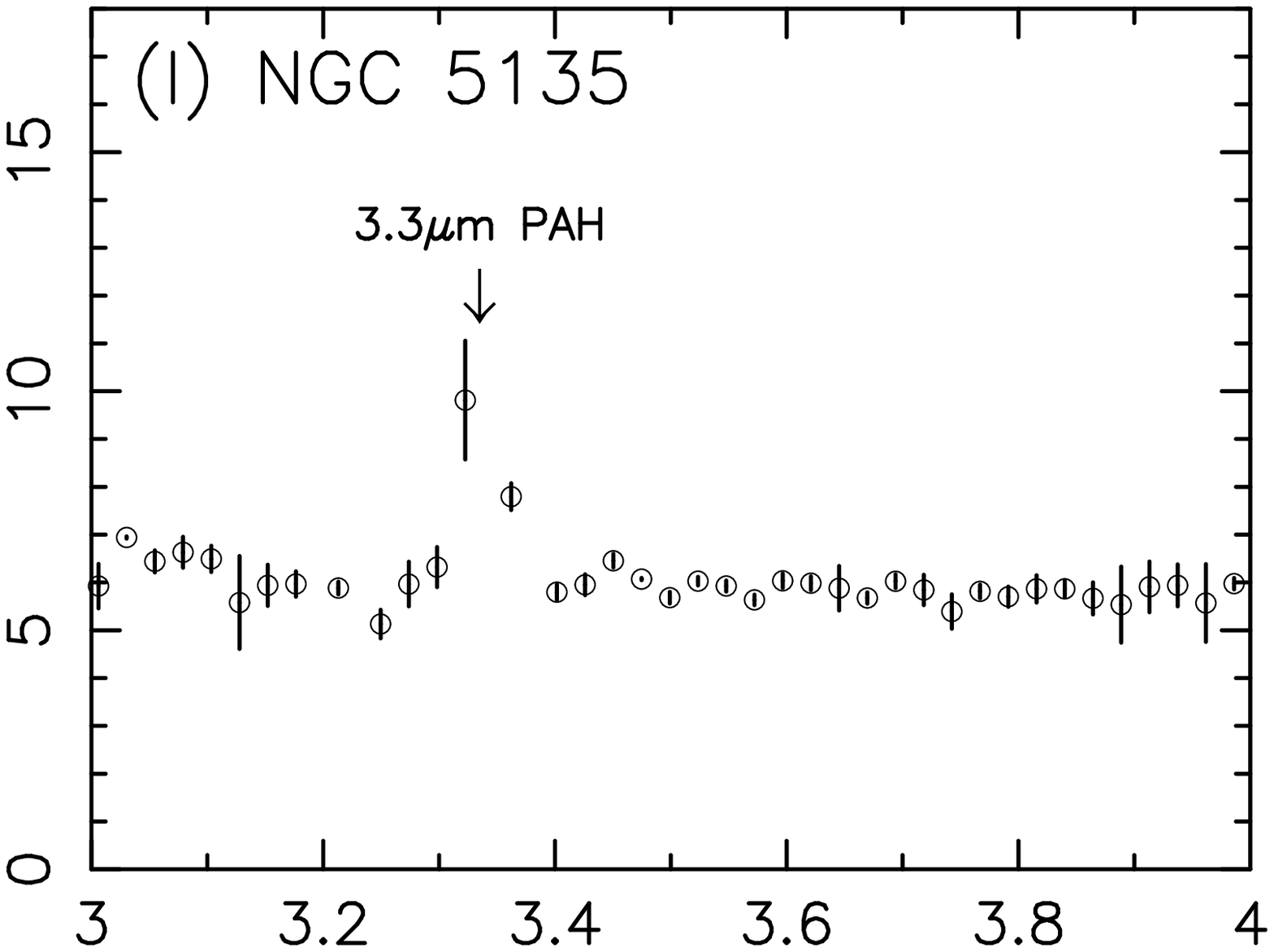}
\end{figure}

\clearpage

\begin{figure}
\plotone{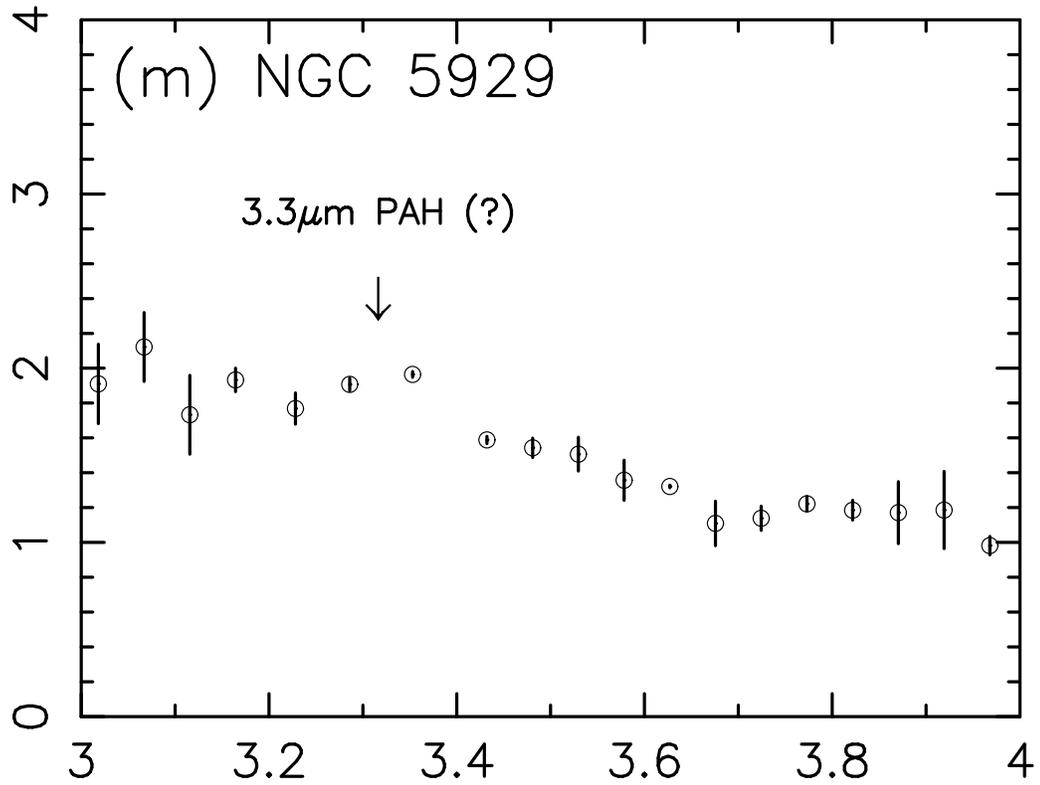}
\caption{3--4 $\mu$m spectra of the 13 Seyfert 2 nuclei.
The abscissa and ordinate are the observed wavelength in $\mu$m and
F$_{\lambda}$ in 10$^{-15}$ W m$^{-2}$ $\mu$m$^{-1}$, respectively.
For Mrk 78 and Mrk 477, the dotted lines are the best fit for the 3.3
$\mu$m PAH emission feature.  
For Mrk 463 and NGC 1068, the dotted lines are the adopted continuum
levels, with respect to which the optical depths of the 3.4 $\mu$m dust
absorption are measured.
\label{fig1}}
\end{figure}

\clearpage

\begin{figure}
\plotone{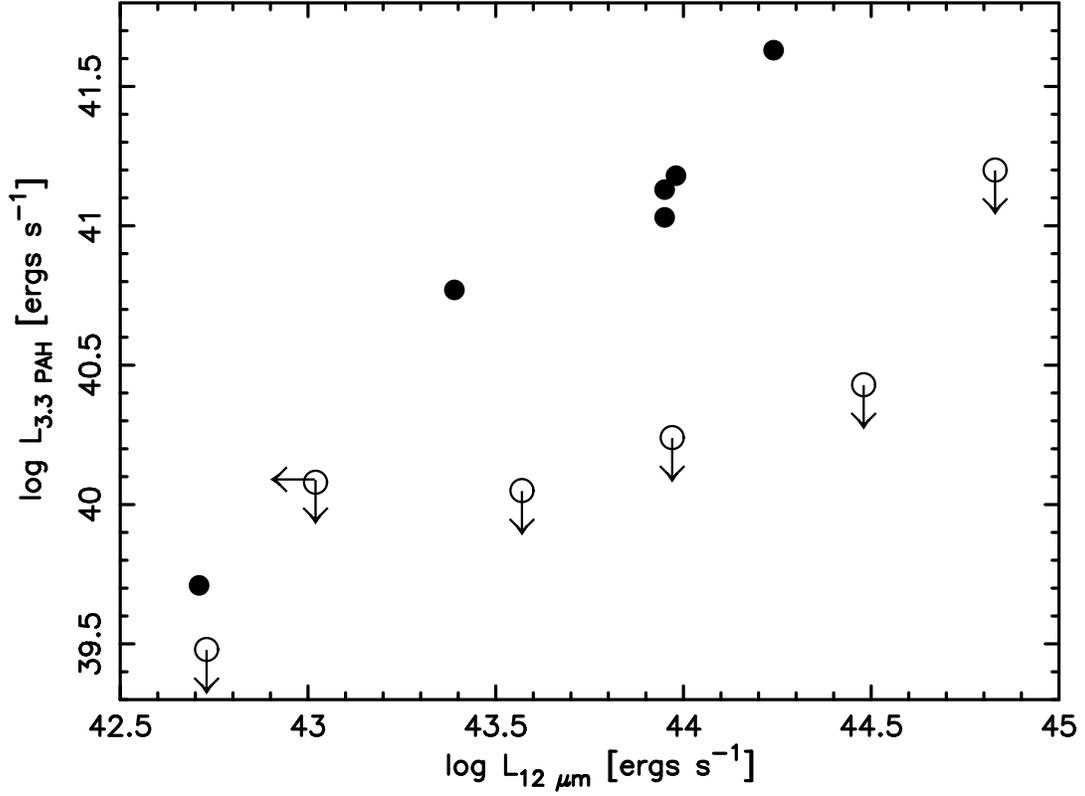}
\caption{
The relation between the logarithm of the 12 $\mu$m luminosity
($\nu$F$_{\nu}$) in ergs s$^{-1}$  (abscissa) and the 3.3 $\mu$m PAH
emission luminosity in ergs s$^{-1}$ measured with our slit spectra
(ordinate).
The filled and open circles are Seyfert 2 nuclei with detected and
non-detected 3.3 $\mu$m PAH emission, respectively.
NGC 5033 is excluded because no meaningful upper limit for the 3.3 $\mu$m
PAH emission is given.
\label{fig2}}
\end{figure}


\clearpage

\begin{thebibliography}{}
\bibitem[Alonso-Herrero et al.(1998)]{alo98}
         Alonso-Herrero, A., Simpson, C., Ward, M. J., \& Wilson, A. S.
         1998, ApJ, 495, 196
\bibitem[Alonso-Herrero et al.(2001)]{alo01}
          Alonso-Herrero, A., Quillen, A. C., Simpson, C., Efstathiou, A.
          \& Ward, M. J. 2001, AJ, 121, 1369
\bibitem[Antonucci(1993)]{ant93}
         Antonucci, R. 1993, ARA\&A, 31, 473
\bibitem[Baker et al.(2001)]{bak01}
         Baker, A. J., Lutz, D., Genzel, R., Tacconi, L. J., \&
         Lehnert, M. D. 2001, A\&A, 372, L37
\bibitem[Calzetti et al.(2000)]{cal00}
         Calzetti, D., Armus, L., Bohlin, R. C., Kinney, A. L.,
         Koornneef, J., \& Storchi-Bergmann, T. 2000, ApJ, 533, 682
\bibitem[Cid Fernandes \& Terlevich (1995)]{cid95}
         Cid Fernandes, R. Jr., \& Terlevich, R. 1995, MNRAS, 272, 423
\bibitem[Clavel et al.(2000)]{cla00}
         Clavel, J. et al. 2000, A\&A, 357, 839
\bibitem[Dudley(1999)]{dud99}
         Dudley, C. C., 1999, MNRAS, 307, 553
\bibitem[Fischer(2000)]{fis00}
         Fischer, J. 2000, in ISO Beyond the Peaks, ed. A. Salama,
         M. F. Kessler, K., Leech, \& B. Schulz (ESA SP-456; Noordwijk:
         ESA), 239 (astro-ph/0009395) 
\bibitem[Genzel \& Cesarsky(2000)]{gh00}
         Genzel, R., \& Cesarsky, C. J. 2000, ARA\&A, 38, 761
\bibitem[Glass \& Moorwood(1985)]{gla85}
         Glass, I. S., \& Moorwood, A. F. M. 1985, MNRAS, 214, 429
\bibitem[Gonzalez Delgado et al.(1998)]{gon98}
         Gonzalez Delgado, R. M., Heckman, T., Leitherer, C.,
         Meurer, G., Krolik, J., Wilson, A. S., Kinney, A., \&
         Koratkar, A. 1998, ApJ, 505, 174
\bibitem[Gonzalez Delgado et al.(2001)]{gon01}
         Gonzalez Delgado, R. M., Heckman, T., \& Leitherer, C.
         2001, ApJ, 546, 845
\bibitem[Goodrich, Veilleux, \& Hill(1994)]{good94}
         Goodrich, R. W., Veilleux, S., \& Hill, G. J. 1994, ApJ,
         422, 521
\bibitem[Heckman(1999)]{hec99}
         Heckman, T. M. 1999, astro-ph/9912029
\bibitem[Heckman(2000)]{hec00}
         Heckman, T. M. 2000, astro-ph/0009075
\bibitem[Heckman et al.(1997)]{hec97}
         Heckman, T. M., Gonzalez Delgado, R., Leitherer, C., Meurer,
         G. R., Krolik, J., Wilson, A. S., Koratkar, A., \&
         Kinney, A. 1997, ApJ, 482, 114
\bibitem[Helou et al.(2000)]{hel00}
         Helou, G., Lu, N. Y., Werner, M. W., Malhotra, S., \&
         Silbermann, N. 2000, ApJ, 532, L21
\bibitem[Imanishi(2000)]{ima00}
         Imanishi, M. 2000, MNRAS, 319, 331
\bibitem[Imanishi(2001)]{ima01}
         Imanishi, M. 2001, AJ, 121, 1927
\bibitem[Imanishi \& Dudley(2000)]{imd00}
         Imanishi, M., \& Dudley, C. C. 2000, ApJ, 545, 701
\bibitem[Imanishi et al.(1997)]{ima97}
         Imanishi, M., Terada, H., Sugiyama, K., Motohara, K.,
         Goto, M., \& Maihara, T. 1997, PASJ, 49, 69
\bibitem[Isobe, Feigelson, \& Nelson(1986)]{iso86}
         Isobe, T., Feigelson, E. D., \& Nelson, P. I. 1986, ApJ, 306, 490
\bibitem[Ivanov et al.(2000)]{iva00}
         Ivanov, V. D., Rieke, G. H., Groppi, C. E.,
         Alonso-Herrero, A., Rieke, M. J., \& Engelbracht, C. W.
         2000, ApJ, 545, 190
\bibitem[Kennicutt(1998)]{ken98}
         Kennicutt, R. 1998, ApJ, 498, 541
\bibitem[Laureijs et al.(2000)]{lau00}
         Laureijs, R. J. et al. 2000, A\&A, 359, 900
\bibitem[Lawrence et al.(1985)]{law85}
         Lawrence, A., Ward, M., Elvis, M., Fabbiano, G.,
         Willner, S. P., Carleton, N. P., \& Longmore, A.
         1985, ApJ, 291, 117
\bibitem[Le Floc'h et al.(2001)]{lef01}
         Le Floc'h, E., Mirabel, I. F., Laurent, O., Charmandaris, V.,
         Gallais, P., Sauvage, M., Vigroux, L., \& Cesarsky, C. 2001,
         A\&A, 367, 487
\bibitem[Levenson, Weaver, \& Heckman(2001)]{lev01}
         Levenson, N. A., Weaver, K. A., \& Heckman, T. M.
         2001, ApJ, 550, 230
\bibitem[Lutz et al.(1996)]{lut96}
         Lutz, D. et al. 1996, A\&A, 315, L269
\bibitem[Marco \& Alloin(2000)]{mar00}
         Marco, O., \& Alloin, D. 2000, A\&A, 353, 465
\bibitem[Mazzarella \& Boroson(1993)]{maz93}
         Mazzarella, J. M., \& Boroson, T. A. 1993, ApJS, 85, 27
\bibitem[Mazzarella et al.(1991)]{maz91}
         Mazzarella, J. M., Gaume, R. A., Soifer, B. T., Graham,
         J. R., Neugebauer, G., \& Matthews, K. 1991, AJ, 102, 1241
\bibitem[Mennella et al.(2001)]{men01}
         Mennella, V., Munoz Caro, G. M., Ruiterkamp, R., Schutte, W. A., 
         Greenberg, J. M., Brucato, J. R., \& Colangeli, L. 2001, A\&A, 
         367, 355
\bibitem[Moorwood(1986)]{moo86}
         Moorwood, A. F. M. 1986, A\&A, 166, 4
\bibitem[Mountain et al.(1990)]{mou90}
         Mountain, C. M., Robertson, D. J., Lee, T. J., \& Wade, R.
         1990, Proc. SPIE, 1235, 25
\bibitem[Mouri et al.(1990)]{mouri90}
         Mouri, H., Kawara, K., Taniguchi, Y., \& Nishida, M.
         1990, ApJ, 356, L39
\bibitem[Murayama et al.(2000)]{mura00}
         Murayama, T., Mouri, H., \& Taniguchi, Y. 2000, ApJ, 528, 179
\bibitem[Pendleton et al.(1994)]{pen94}
         Pendleton, Y. J., Sandford, S. A., Allamandola, L. J.,
         Tielens, A. G. G. M., \& Sellgren, K. 1994, ApJ, 437, 683
\bibitem[Rieke(1978)]{rie78}
         Rieke, G. H. 1978, ApJ, 226, 550
\bibitem[Rieke \& Lebofsky(1985)]{rie85}
         Rieke, G. H., \& Lebofsky, M. J. 1985, ApJ, 288, 618
\bibitem[Sanders \& Mirabel(1996)]{sam96}
         Sanders, D. B., \& Mirabel, I. F. 1996, ARA\&A, 34, 749
\bibitem[Savage \& Mathis(1979)]{sav79}
         Savage, B. D., \& Mathis, J. S. 1979, ARA\&A, 17, 73
\bibitem[Scoville et al.(2000)]{sco00}
         Scoville, N. Z. et al. 2000, AJ, 119, 991
\bibitem[Shure et al.(1994)]{shu94}
         Shure, M. A., Toomey, D. W., Rayner, J. T., Onaka, P., \&
         Denault, A. J. 1994, Proc. SPIE, 2198, 614
\bibitem[Simpson(1998)]{sim98}
         Simpson, C. 1998, ApJ, 509, 653
\bibitem[Smith et al.(1989)]{smi89}
         Smith, R. G., Sellgren, K., \& Tokunaga, A. T. 1989, ApJ, 344, 
         413
\bibitem[Spoon et al.(2000)]{spo00}
         Spoon, H. W. W., Koornneef, J., Moorwood, A. F. M., Lutz, D., 
         \& Tielens, A. G. G. M. 2000, A\&A, 357, 898
\bibitem[Storchi-Bergmann et al.(2000)]{sto00}
         Storchi-Bergmann, T., Raimann, D., Bica, E. L. D., \&
         Fraquelli, H. A. 2000, ApJ, 544, 747
\bibitem[Surace \& Sanders(1999)]{sur99}
         Surace, J. A., \& Sanders, D. B. 1999, ApJ, 512, 162
\bibitem[Surace, Sanders, \& Evans(2000)]{sur00}
         Surace, J. A., Sanders, D. B., \& Evans, A. S. 2000, ApJ, 529,
         170
\bibitem[Tokunaga et al.(1991)]{tok91}
         Tokunaga, A. T., Sellgren, K., Smith, R. G., Nagata, T., 
         Sakata, A., \& Nakada, Y. 1991, ApJ, 380, 452   
\bibitem[Tokunaga(2000)]{tok00}
         Tokunaga, A. T. 2000, in Allen's
         Astrophysical Quantities, ed.
         A. N. Cox (4th ed: AIP Press: Springer), Chapter 7, p.143
\bibitem[Veilleux, Goodrich, \& Hill(1997)]{vei97}
         Veilleux, S., Goodrich, R. W., \& Hill, G. J. 1997, ApJ, 477,
         631
\bibitem[Voit(1992)]{voit92}
         Voit, G. M. 1992, MNRAS, 258, 841
\bibitem[Zhou, Wynn-Williams, \& Sanders(1993)]{zho93}
         Zhou, S., Wynn-Williams, C. G., \& Sanders, D. B. 1993,
         ApJ, 409, 149
\end{thebibliography}
\end{document}